\documentclass[aps,prd,twocolumn,twoside,preprintnumbers,superscriptaddress,nofootinbib,floatfix]{revtex4-1}
\usepackage[T1]{fontenc}

\usepackage{soul}
\usepackage{cancel}

\usepackage{amsmath,slashed}
\usepackage{graphicx,graphics,color}
\definecolor{darkgreen}{rgb}{0.0, 0.6, 0.0}
\usepackage{dcolumn}
\usepackage{hyperref}
\usepackage{xspace}
\usepackage{enumerate}
\usepackage{varwidth}
\usepackage{physics}
\usepackage{amssymb}
\usepackage{mathtools}
\usepackage{dsfont}
\usepackage{booktabs,tabulary,longtable, multirow}
\hypersetup{
    colorlinks=true,
    linkcolor=blue,
    urlcolor=blue,
    citecolor=blue
}
\newcommand{\mpi}{M_\pi}
\newcommand{\mpic}{M_{\pi^+}}
\newcommand{\mpin}{M_{\pi^0}}
\newcommand{\beq}{\begin{equation}}
\newcommand{\eeq}{\end{equation}}
\newcommand{\diff}{\text{d}}

\newcommand{\Order}{\mathcal{O}}

\newcommand{\GeV}{\,\text{GeV}}
\newcommand{\MeV}{\,\text{MeV}}

\renewcommand{\Im}{\text{Im}}

\allowdisplaybreaks[1]

\begin{document}

\title{\texorpdfstring{Dispersive analysis of the $\boldsymbol{J/\psi \to \gamma \pi^0 \pi^0}$ process}{Dispersive analysis of the J/psi -> gamma pi0 pi0 process}}
\author{Bai-Long Hoid}
\affiliation{Institut f\"ur Kernphysik and PRISMA$^{++}$  Cluster of Excellence, Johannes Gutenberg-Universit\"at Mainz,  55099 Mainz, Germany}
\author{Igor Danilkin}
\affiliation{Institut f\"ur Kernphysik and PRISMA$^{++}$  Cluster of Excellence, Johannes Gutenberg-Universit\"at Mainz,  55099 Mainz, Germany}
\author{Anna Testa}
\affiliation{Institut f\"ur Kernphysik and PRISMA$^{++}$  Cluster of Excellence, Johannes Gutenberg-Universit\"at Mainz,  55099 Mainz, Germany}
\affiliation{Istituto Nazionale di Fisica Nucleare, Sezione di Padova, 35131 Padova, Italy}
\affiliation{Dipartimento di Fisica e Astronomia ``Galileo Galilei'', Universit\`a di Padova, 35131 Padova, Italy}
\author{Marc Vanderhaeghen}
\affiliation{Institut f\"ur Kernphysik and PRISMA$^{++}$  Cluster of Excellence, Johannes Gutenberg-Universit\"at Mainz,  55099 Mainz, Germany}

\begin{abstract}
We present a dispersive amplitude analysis of the low-energy $\pi^0\pi^0$ system in the radiative decay $J/\psi\to\gamma\pi^0\pi^0$, using the mass-independent BESIII $0^{++}$ and $2^{++}$ intensities, the total spectrum, and the measured $0^{++}-2^{++}$ $E1$ phase difference. The isoscalar $S$-wave is described by a coupled-channel $\pi\pi/K\bar K$ Muskhelishvili--Omn\`es  representation, which implements the strong final-state interactions associated with the $f_0(500)$ and $f_0(980)$.  The $D$-wave is treated with a single-channel $\pi\pi$ Muskhelishvili--Omn\`es representation, where we identify all kinematic constraints of helicity amplitudes before transforming them to the experimental $E1$, $M2$, and $E3$ multipoles. Smooth short-distance production of a pseudoscalar-meson pair is encoded in subtraction polynomials, while left-hand-cut effects are estimated and found to be numerically subleading. We identify the negative solution of the BESIII $0^{++}-2^{++}$ $E1$ phase ambiguity, after using the modulo-$\pi$ freedom of production amplitudes, as the phase solution compatible with unitarity constraints, showing that the measured phase information can be accommodated with the Omn\`es phase motion without requiring large additional phases. By normalizing the BESIII intensities with the extracted branching fraction, we fix the absolute scale of the fitted amplitudes, making them suitable as input for future dispersive studies of two-pion contributions to gravitational form factors.
\end{abstract}

\maketitle
\section{Introduction}

\begin{figure*}[t]
	\centering
	\includegraphics[width=0.78\textwidth]{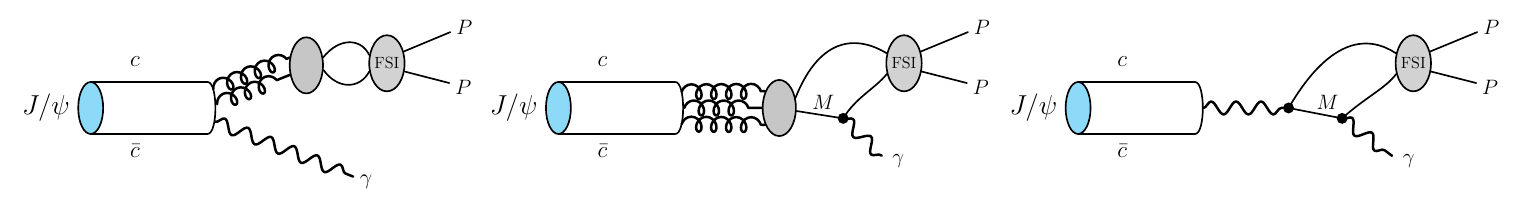}
	\caption{Main decay mechanisms for $J/\psi\to\gamma P P$, where $P=\pi,K$ and the corresponding charge-conjugate states are implied. From left to right: dominant direct $c\bar c\to\gamma gg$ production followed by hadronization into a pseudoscalar-meson pair; and subleading meson-exchange topologies, in which $P$ and an intermediate meson $M$ are produced through gluonic or electromagnetic mechanisms, followed by the transition $M\to\gamma P$.  These exchange topologies are expected to be small compared with the direct radiative production mechanism; in the BESIII $\pi^0\pi^0$ analysis~\cite{BESIII:2015rug}, events with any $\gamma\pi^0$ invariant mass within $50\,\text{MeV}$ of the $\omega$ mass were vetoed to  further suppress the $J/\psi \to \omega\pi^0\to \gamma\pi^0\pi^0$ background.
    }
	\label{fig:Feyn}
\end{figure*}

Radiative decays of the $J/\psi$, such as  $J/\psi\to\gamma\pi^0\pi^0$ and $J/\psi\to\gamma K_SK_S$~\cite{BESIII:2015rug,BESIII:2018ubj},  provide a useful laboratory for light-meson spectroscopy. They further offer a promising environment for searches for QCD exotics that are difficult to explain within a simple quark–antiquark picture~\cite{Kopke:1988cs,Klempt:2007cp}. Owing to the Okubo--Zweig--Iizuka~\cite{Okubo:1963fa,Zweig:1964jf,Iizuka:1966fk} (OZI) and electromagnetic suppressions, radiative $J/\psi$ decays proceed at short distances predominantly via $c\bar c\to\gamma gg$, as illustrated in Fig.~\ref{fig:Feyn}. This makes radiative $J/\psi\to\gamma+\text{hadrons}$ processes sensitive to scalar and tensor resonances and, in particular, to states with an enhanced gluonic component such as glueball candidates~\cite{Bali:1993fb,Morningstar:1999rf,Chen:2005mg,Gregory:2012hu,Gui:2012gx}. The low-energy $J/\psi\to\gamma\pi^0\pi^0$ amplitudes are also relevant for pion and proton gravitational form factors, which are encoded in matrix elements of the QCD energy--momentum tensor.  Near-threshold $J/\psi$ photoproduction has been used to access the proton's gluonic gravitational form factors~\cite{Kharzeev:2021qkd,Duran:2022xag}.  In $t$-channel dispersive representations, two-pion intermediate states enter the nucleon $D$-term and related gravitational form factors at low momentum transfers~\cite{Pasquini:2014vua,Cao:2024zlf}; applying this strategy to $J/\psi$ photoproduction requires reliable $0^{++}$ and $2^{++}$ $J/\psi\to\gamma\pi\pi$ amplitudes as hadronic input. These applications motivate improved experimental measurements and better theoretical control of the scalar and tensor meson sectors, especially their final-state interactions. 

On the experimental side, BESIII conducted the most precise measurement of $J/\psi\to\gamma \pi^0 \pi^0$ based on a high-statistics data sample of $1.311(11)\times10^{9}$ $J/\psi$ decays~\cite{BESIII:2015rug}. Compared to an earlier partial-wave analysis limited to Breit--Wigner assumptions and a $\pi\pi$ mass range above $1\GeV$~\cite{BES:2006ssn}, the BESIII study performed a binned mass-independent amplitude analysis for the scalar and tensor components of the $ \pi^0 \pi^0$ system. This avoids imposing a fixed set of resonant poles from the beginning, which is especially important in the scalar sector, where broad and overlapping states make the identification of individual resonances challenging. In this way, the BESIII result provides model-independent experimental constraints on the scalar $0^{++}$ and tensor $2^{++}$ contributions in $J/\psi\to\gamma \pi^0 \pi^0$, and is therefore particularly valuable as input for global and systematic amplitude analyses. These data have already been applied in coupled-channel analyses to identify the scalar and tensor resonance spectrum~\cite{Sarantsev:2021ein,Rodas:2022jpsi}.

The underlying mechanisms for $J/\psi\to\gamma\pi^0\pi^0$, as shown in Fig.~\ref{fig:Feyn}, provide a useful way to organize theoretical discussions of scalar-meson production dynamics and final-state interactions in the system: the dominant direct mechanism is associated with production terms dressed by final-state interactions, while the OZI- or electromagnetically suppressed meson-exchange topologies generate subleading left-hand-cut (LHC) contributions.  This structure is partly reflected in previous theoretical studies: Ref.~\cite{Xiao:2019lrj} investigated radiative $J/\psi$ decays including coupled-channel final-state interactions, emphasizing the role of the $K\bar K$ channel in the $f_0(980)$ contribution to $J/\psi\to\gamma\pi\pi$; Ref.~\cite{Sakai:2019uig} included $J/\psi\to\gamma\pi^+\pi^-$ within a vector-meson-dominance framework, relating hadronic and radiative $J/\psi$ decays to study the $f_0(500)$ and $f_0(980)$ regions. The production dynamics considered in these works do not aim to describe the BESIII $J/\psi\to\gamma\pi^0\pi^0$ data~\cite{BESIII:2015rug}, and they predict rather small branching ratios compared with the observed strengths. 

More recently, the scalar $0^{++}$ intensity of the BESIII $J/\psi\to\gamma\pi^0\pi^0$ data was analyzed in Ref.~\cite{Achasov:2020aun} in connection with the nature of the light scalar mesons. Reference~\cite{Danilkin:2025kyo} constructed a coupled-channel Omn\`es matrix for the isoscalar $D$-wave $\pi\pi/K\bar K$ system, incorporating the $f_2(1270)$ and $f_2'(1525)$ resonances in a framework consistent with analyticity and unitarity. It was then applied to a simultaneous description of the BESIII spectra~\cite{BESIII:2015rug,BESIII:2018ubj} from the dominant $2^{++}$ $E1$ amplitudes. However, the precise total-intensity spectrum, the subleading $2^{++}$ multipoles, and the measured $0^{++}-2^{++}$ $E1$ phase difference have not yet been fully exploited, and can provide additional constraints on the production dynamics and final-state interactions.

Dispersive methods provide a controlled implementation of analyticity and coupled-channel unitarity.  These constraints are key for the strong $\pi\pi/K\bar K$ final-state interactions in the isoscalar-scalar channel, where the broad $f_0(500)$ and the rapid structure around the $f_0(980)$ dominate the low-energy region. Different types of Muskhelishvili--Omn\`es (MO) representations~\cite{Garcia-Martin:2010kyn,Moussallam:2013una} have been investigated in detail for  $\gamma^*\gamma^{(*)}\to\pi\pi$~\cite{Danilkin:2018qfn,Hoferichter:2019nlq,Danilkin:2019opj}, and the resulting helicity partial waves provide an essential ingredient for dispersive evaluations of the two-pion contribution to hadronic light-by-light scattering in the muon $g-2$~\cite{Colangelo:2017fiz,Colangelo:2017qdm,Danilkin:2021icn}. Radiative decays of the $\phi$ meson provide a closely related time-like application, in which $\pi\pi/K\bar K$ rescattering is combined with LHCs from the Born term and vector-meson exchanges~\cite{Hoid:2026atr}. In this work we use a coupled-channel MO representation for the $S$-wave final-state interactions up to $1.05\GeV$, while the $D$-wave is described by a single-channel $\pi\pi$ MO representation up to $1.45\GeV$. The dominant production strength is encoded in subtraction polynomials. Possible LHC effects from Born terms and spin-one exchanges are estimated separately and found to be subleading. We incorporate all three measured $2^{++}$ multipoles and the precise total-intensity spectrum that is free from partial-wave ambiguities. We also address, for the first time,  the phenomenological impact of the presented $0^{++}-2^{++}$ $E1$ phase difference, whose conventional positive solution deviates strongly from the elastic Watson-theorem expectation. By normalizing the BESIII intensities with the extracted branching fraction, we fix the absolute scale of the fitted amplitudes, making them suitable as input for future dispersive studies of two-pion contributions to gravitational form factors.

The paper is organized as follows.  In Sect.~\ref{sec:form}, we introduce the kinematics, helicity amplitudes,  and their partial-wave expansion. The corresponding kinematic constraints and the multipole amplitudes are introduced thereafter. We show the dispersive representations of the $S$- and $D$-waves in Sect.~\ref{sec:disp-reps}.  The fit strategy and comparison with the $J/\psi\to\gamma\pi^0\pi^0$ data are discussed in Sect.~\ref{sec:results}.  Conclusions and possible applications are given in Sect.~\ref{sec:conc}. 

\section{Formalism}
\label{sec:form}

\subsection{Kinematics and helicity amplitudes}
We consider the radiative decay
\begin{equation}
	J/\psi(q)\to \gamma(q_1)\pi^0(p_1)\pi^0(p_2)\, ,
\end{equation}
with 
\begin{equation}
	q^2=M_{J/\psi}^2\, , \qquad q_1^2=0\, , \qquad s=(p_1+p_2)^2\, .
\end{equation}
The remaining Mandelstam variables are
\begin{equation}
	t=(q_1+p_1)^2\, ,\qquad u=(q_1+p_2)^2\, ,
\end{equation}
which obey 
\begin{equation}
	s+t+u=q^2+2\mpin^2\, .
\end{equation}
For the $\pi^0\pi^0$ final state, Bose symmetry requires the amplitude to be invariant under $p_1 \leftrightarrow p_2$, or equivalently $t \leftrightarrow u$. The helicity angle $\theta$ is defined in the two-pion center-of-mass frame as the angle between the outgoing photon and  the pion with momentum $p_1$.

Our analysis is based on the helicity formalism applied to the radiative decays of vector mesons~\cite{Moussallam:2021dpk,Hoid:2026atr}.  To isolate the analytic structure relevant for the dispersive treatment, we decompose the hadronic tensor $H^{\mu\nu}$ into gauge-invariant Lorentz structures and scalar invariant amplitudes free of kinematic singularities and constraints~\cite{Bardeen:1968ebo,Tarrach:1975tu,Drechsel:1997xv},   
\begin{equation}
	H^{\mu\nu}=\sum_{i=1}^{3}F_i(s,t)\,L_i^{\mu\nu}\, ,
\end{equation}
where the tensors $L_i^{\mu\nu}$ form a complete basis for the process~\cite{Danilkin:2018qfn}. Choosing the helicities of the photon and the $J/\psi$ as $\lambda_{1}$ and $\lambda_{2}$, the helicity amplitudes are obtained by contracting $H^{\mu\nu}$ with the polarization vectors, 
\begin{equation}
	e_\gamma^{*\mu}(q_1,\lambda_1)e_\psi^\nu(q,\lambda_2)H_{\mu\nu}\equiv e^{i(\lambda_2-\lambda_1)\varphi}H_{\lambda_1\lambda_2}(s,\theta)\, .
\end{equation}
Parity leaves three independent helicity amplitudes, which we choose as $H_{++}$, $H_{+-}$, and $H_{+0}$. 

The isospin decomposition of the $\pi^0\pi^0$ channel is tied to the dominant production mechanism.  The short-distance source $c\bar c\to\gamma gg$ produces an isoscalar two-pion system, while isospin-violating contributions in $J/\psi$ decays are expected to be dominated by electromagnetic transitions~\cite{Chen:2014yta,BaldiniFerroli:2016mbs,Cao:2025ncx}.\footnote{See also the discussion in Ref.~\cite{Rodas:2022jpsi}.}  The main isospin-violating process $J/\psi\to\omega\pi^0\to\gamma\pi^0\pi^0$ is further reduced by  the BESIII event selection~\cite{BESIII:2015rug}.  We therefore assume isospin conservation, which leads to
\begin{equation}
	H_{\lambda_1\lambda_2}^{0}
	=-\sqrt{3}\,H_{\lambda_1\lambda_2}^{n}
	=-\sqrt{3}\,H_{\lambda_1\lambda_2}^{c}\, ,
\end{equation}
where $n$ and $c$ denote the neutral- and charged-pion channels. The isoscalar kaon component in the coupled-channel $S$-wave is defined analogously by
\begin{equation}
	K_{\lambda_1\lambda_2}^{0}
	=-\frac{1}{\sqrt{2}}\left(
	K_{\lambda_1\lambda_2}^{c}
	+K_{\lambda_1\lambda_2}^{n}\right) .
\end{equation}

\subsection{Partial waves and kinematic constraints}
The helicity amplitudes are expanded in partial waves as~\cite{Jacob:1959at}
\begin{align}
	H_{\lambda_1\lambda_2}^{0}(s,\theta)
	&=\sum_J (2J+1)\,h^0_{J,\lambda_1\lambda_2}(s)\,
	d^J_{\lambda_2-\lambda_1,0}(\theta)\, ,\notag\\
	K_{\lambda_1\lambda_2}^{0}(s,\theta)
	&=\frac{1}{\sqrt{2}}\sum_J (2J+1)\,k^0_{J,\lambda_1\lambda_2}(s)\,
	d^J_{\lambda_2-\lambda_1,0}(\theta)\, ,
\end{align}
where $d^J_{\lambda,\bar{\lambda}}(\theta)$ denotes a Wigner rotation function. The extra factor $1/\sqrt{2}$ in the $K\bar K$ expansion follows the convention of Ref.~\cite{Garcia-Martin:2010kyn}, so that the coupled-channel partial waves obey the same unitarity normalization for the $\pi\pi$ and $K\bar K$ channels. Due to Bose symmetry of the $\pi^0\pi^0$ state, only even values of $J$ contribute to the expansion: $0^{++}$ and $2^{++}$ contributions are therefore dominant, while the $4^{++}$ amplitude is found to be insignificant~\cite{BESIII:2015rug}. 

Helicity partial-wave amplitudes $h^I_{J,\lambda_1\lambda_2}$ and $k^I_{J,\lambda_1\lambda_2}$ can contain kinematic singularities and satisfy nontrivial relations at thresholds and pseudothresholds. These constraints can be imposed either in terms of projected invariant amplitudes or diagonalized Roy--Steiner kernels for the Born-subtracted amplitudes~\cite{Danilkin:2018qfn,Hoferichter:2019nlq}. Since the charged-kaon and charged-pion Born terms are numerically negligible in the present application (see App.~\ref{app:LHCs}), the constraints below are written directly for the amplitudes rather than for Born-subtracted quantities. For the $S$-wave, Low's theorem~\cite{Low:1958sn} then requires  
\begin{equation}
	\boldsymbol{h}^{\,0}_{0,++}(s)
	\equiv \begin{pmatrix} h^0_{0,++}(s)\\  k^0_{0,++}(s)\end{pmatrix}
	=(s-q^2)\,\hat{\boldsymbol{h}}^0_{0,++}(s)\, ,
	\label{eq:swave-soft-photon}
\end{equation}
where $\hat{\boldsymbol{h}}^0_{0,++}$ is regular at the soft-photon point $s=q^2$.

For the $D$-wave, partial-wave amplitudes obey a hierarchy of threshold and soft-photon constraints.  In the present analysis these constraints are applied to the single-channel $\pi\pi$ $D$-wave amplitudes. Following the procedure of Ref.~\cite{Danilkin:2018qfn} for $\gamma\gamma^{*}\to\pi\pi$, we adapt the construction to the decay kinematics of $J/\psi\to\gamma\pi\pi$,
	\begin{align}
	h^0_{2,+-}(s)
	&=(s-4\mpi^2)(s-q^2)h^0_{2,1}(s)\, ,\notag\\
	h^0_{2,+-}(s)-\sqrt{\frac{2s}{q^2}}h^0_{2,+0}(s)
	&=(s-4\mpi^2)(s-q^2)^2h^0_{2,2}(s)\, ,\notag\\
    h^0_{2,+-}(s)-2\sqrt{\frac{2s}{q^2}}h^0_{2,+0}(s)
    &+\frac{\sqrt{6}\,s}{q^2}h^0_{2,++}(s)\notag\\
    &=(s-4\mpi^2)(s-q^2)^3h^0_{2,3}(s)\, .
    \label{eq:dwave-helicity-constraints}
\end{align}
where we choose $\mpi=\mpic$ in the isospin limit. The functions $h^0_{2,i}$ are regular at both the two-pion threshold and the soft-photon point. 

The kinematic constraints in Eqs.~\eqref{eq:swave-soft-photon} and \eqref{eq:dwave-helicity-constraints} are imposed before introducing dispersive representations, so that $\hat{\boldsymbol{h}}^0_{0,++}$ and $h^0_{2,i}$ encode only the remaining dynamical information.

\subsection{Multipole amplitudes}
Radiative decays of a charmonium state into a hadronic state $X$ with definite spin and parity can be described equivalently in a helicity basis or in a radiative multipole basis.  The helicity basis is natural for partial-wave projection and for separating the kinematic constraints discussed above.  The radiative multipole basis instead classifies the same transition by the angular momentum carried by the emitted photon.  It is therefore well suited for testing the hierarchy of electromagnetic multipoles in experiments. The BESIII mass-independent analysis of $J/\psi\to\gamma\pi^0\pi^0$ reports the $0^{++}$ and $2^{++}$ components in this basis~\cite{BESIII:2015rug}.

For $J/\psi(1^{--})\to\gamma X$, with $X$ of spin $J_X$, angular-momentum conservation gives $|1-J_X|\leq J_\gamma\leq 1+J_X$, and parity conservation determines whether the multipole transition with angular momentum $J_\gamma$ is electric or magnetic.  Thus a $0^{++}$ two-pion state admits only an $E1$ transition, whereas a $2^{++}$ state has the three allowed transitions $E1$, $M2$, and $E3$. Here $E1$, $M2$, and $E3$ denote electric-dipole, magnetic-quadrupole, and electric-octupole radiative multipoles.   

We use the basis transformation employed in charmonium radiative transitions~\cite{Karl:1975qf,Olsson:1986dn,Sebastian:1992qe}.  For the amplitudes considered here, the helicity partial waves with $\lambda_1=+1$ are related to the multipole amplitudes by
\begin{align}
	h^0_{J,+\lambda_2}(s)
	&=\sum_{J_\gamma=1}^{J+1}
	\sqrt{\frac{2J_\gamma+1}{2J+1}}\,
	\bigl\langle J_\gamma,1;\,1,\nu-1 \bigm| J,\nu\bigr\rangle
	\notag\\
	&\quad\times h^0_{J,\mu_\gamma(J_\gamma)}(s),
	\qquad \nu=|1-\lambda_2|,\notag\\
	\mu_\gamma(J_\gamma)&=
	\begin{cases}
		EJ_\gamma,& J_\gamma\ \text{odd},\\
		MJ_\gamma,& J_\gamma\ \text{even}.
	\end{cases}
\end{align}
Here the bracket denotes a Clebsch--Gordan coefficient, and the multipole labels refer to the radiative transition.

The transformation is trivial for the $S$-wave: the $J=0$ partial wave is associated with the $0^{++}$ $E1$ amplitude,
\begin{equation}
	 h^0_{0,++}(s)=h^0_{0,E1}(s)\, . 
\end{equation}
For the physical neutral-pion channel this gives
\begin{equation}
	h^{n}_{0,E1}(s)=-\frac{1}{\sqrt{3}}\,h^0_{0,E1}(s)\, .
\end{equation}
Evaluating the Clebsch--Gordan coefficients, we find for the $2^{++}$ components,
\begin{widetext}
\begin{equation}
	\begin{pmatrix} h^0_{2,+-}\\ h^0_{2,+0}\\ h^0_{2,++} \end{pmatrix}=
	\begin{pmatrix} \sqrt{\frac{3}{5}} & -\frac{1}{\sqrt{3}} & \frac{1}{\sqrt{15}}\\ \sqrt{\frac{3}{10}} & \frac{1}{\sqrt{6}} & -2\sqrt{\frac{2}{15}}\\ \frac{1}{\sqrt{10}} & \frac{1}{\sqrt{2}} & \sqrt{\frac{2}{5}} \end{pmatrix}
	\begin{pmatrix} h^0_{2,E1}\\ h^0_{2,M2}\\ h^0_{2,E3} \end{pmatrix}.
	\label{eq:helicity-multipole}
\end{equation}
\end{widetext}
For the  $2^{++}$ radiative multipoles, threshold behavior and angular-momentum factors at the soft-photon point lead to the hierarchy
\begin{align}
	h^0_{2,E1}(s)&=\Order\!\left((s-4M_{\pi}^2)(s-q^2)\right) ,\notag\\
	h^0_{2,M2}(s)&=\Order\!\left((s-4M_{\pi}^2)(s-q^2)^2\right) ,\notag\\
	h^0_{2,E3}(s)&=\Order\!\left((s-4M_{\pi}^2)(s-q^2)^3\right) .
	\label{eq:dwave-multipole-constraints}
\end{align}
The kinematic constraints for the helicity amplitudes in Eq.~\eqref{eq:dwave-helicity-constraints} directly reproduce this behavior in Eq.~\eqref{eq:dwave-multipole-constraints}. We stress that Eq.~\eqref{eq:dwave-multipole-constraints} displays only the dominant soft-photon behavior of the individual multipoles. The full kinematic constraints obtained from the Lorentz-basis analysis are
stronger: they imply the helicity-amplitude relations in Eq.~\eqref{eq:dwave-helicity-constraints},
which translate into correlations among the \(E1\), \(M2\), and \(E3\)
multipoles away from the soft-photon point.

Since Eq.~\eqref{eq:helicity-multipole} is orthogonal, it preserves the angle-integrated $2^{++}$ intensity: it can be written equivalently as an incoherent sum over helicity amplitudes or over multipoles,
\begin{align}
	&\left|h^0_{2,+-}\right|^2+\left|h^0_{2,+0}\right|^2
	+\left|h^0_{2,++}\right|^2 \notag\\
	&\qquad =
	\left|h^0_{2,E1}\right|^2+\left|h^0_{2,M2}\right|^2
	+\left|h^0_{2,E3}\right|^2 .
\end{align}
This transformation is essential for the comparison with BESIII, whose mass-independent solution is reported in the $E1$, $M2$, and $E3$ basis, with the $E1$ component carrying the dominant $2^{++}$ strength. 
For the physical neutral-pion channel, the same isospin factor applies to each $2^{++}$ multipole,
\begin{equation}
	h^{n}_{2,X}(s)=-\frac{1}{\sqrt{3}}\,h^0_{2,X}(s),\qquad X=E1,M2,E3\, .
\end{equation}

\section{Dispersive representations of the $\boldsymbol{S}$- and $\boldsymbol{D}$-waves}
\label{sec:disp-reps}

We now discuss the detailed construction of the dispersive representations used for the $S$- and $D$-waves before fitting the data.  The two cases are presented separately, with the kinematic constraints of Sect.~\ref{sec:form} incorporated into the amplitudes used in the numerical analysis.

\subsection{$\boldsymbol{S}$-wave coupled-channel representation}
The final-state interaction in the isoscalar $S$-wave is governed by coupled $\pi\pi/K\bar K$ dynamics. It is encoded in an Omn\`es matrix $\boldsymbol{\Omega}_0^0(s)$, with its discontinuity fixed by coupled-channel unitarity and the $I=0$, $J=0$ hadronic $T$-matrix. We write
\begin{equation}
	\boldsymbol{\Omega}_0^0(s)=\begin{pmatrix} \Omega_{11}(s) & \Omega_{12}(s)\\ \Omega_{21}(s) & \Omega_{22}(s) \end{pmatrix},
\end{equation}
where channel 1 denotes $\pi\pi$ and channel 2 denotes $K\bar K$.   In the present analysis we use the same two-channel input as in Ref.~\cite{Hoid:2026atr}, based on the data-driven analysis of Ref.~\cite{Danilkin:2020pak}.  The input is constrained by Roy-like analysis of $\pi\pi\to\pi\pi$~\cite{Garcia-Martin:2011iqs} and by dispersive information on $\pi K\to\pi K$ and $\pi\pi\to K\bar K$~\cite{Pelaez:2020gnd}.

In MO representations, the Omn\`es matrix fixes the right-hand-cut structure required by two-channel unitarity.  The remaining input is process dependent and enters through possible LHCs and subtraction parameters.  This framework has been used successfully in radiative $\phi$ decays~\cite{Moussallam:2021dpk,Hoid:2026atr}: once the leading crossed-channel singularities are included, the spectra can be described with a minimal set of subtraction parameters. In decay kinematics, left-hand singularities can in principle overlap with the unitarity cut and thereby modify the simple Watson-theorem phase expectation in the elastic region (see, e.g., $\omega,\phi\to3\pi$ decays~\cite{Niecknig:2012sj, Schneider:2012ez, Danilkin:2014cra, JPAC:2020umo}). Since the BESIII $0^{++}-2^{++}$ $E1$ phase difference shows, at first sight, a strong  deviation from the elastic $\pi\pi$ phase-shift difference~\cite{Rodas:2022jpsi}, we estimate the leading crossed-channel mechanisms explicitly. However, as shown below and in App.~\ref{app:LHCs}, the spin-one-exchange and Born contributions are numerically small in the present process. We therefore use them only as a check of possible interference with the dominant smooth production terms, rather than as the mechanism responsible for the observed spectrum or phase motion.

We then use the standard MO representation, treating the LHC as an inhomogeneous term~\cite{Garcia-Martin:2010kyn,Moussallam:2013una,Hoid:2026atr}.  With the two-component $S$-wave amplitude of Eq.~\eqref{eq:swave-soft-photon} and the vector $\boldsymbol{P}(s)=(P_1(s),P_2(s))^T$ of subtraction polynomials, the $S$-wave amplitude is written as
\begin{align}
	\boldsymbol{h}^{\,0}_{0,++}(s)
	&=\boldsymbol{h}^{\,0,L}_{0,++}(s)\notag\\
	&\quad +(s-q^2)\,\boldsymbol{\Omega}_0^0(s)
	\left(\boldsymbol{P}(s)-\boldsymbol{\Delta}^{\,L}_{0,++}(s;s_0)\right),
\end{align}
where the LHC rescattering correction is evaluated in once-subtracted form,
\begin{align}
	\boldsymbol{\Delta}^{\,L}_{0,++}(s;s_0)
	&=\frac{s-s_0}{\pi}\int_{4M_{\pi}^2}^{\infty}
	\frac{\diff s'}{s'-s_0}\notag\\
	&\quad\times
	\frac{\Im\left(\boldsymbol{\Omega}_0^0(s')\right)^{-1}}
	{(s'-q^2)(s'-s)}
	\boldsymbol{h}^{\,0,L}_{0,++}(s').
\end{align}
Here $s_0$ denotes the subtraction point and for $\boldsymbol{h}^{\,0,L}_{0,++}$ we use the pole parts, while polynomial pieces associated with different off-shell realizations are absorbed into the subtraction polynomials. For the LHC test, we retain only the largest signal-mimicking contribution, the $b_1$-exchange term, which enters through the pion component,
\begin{equation}
	\boldsymbol{h}^{\,0,L}_{0,++}(s)
	=\begin{pmatrix}h^{0,b_1}_{0,++}(s)\\
	0
	\end{pmatrix}.
\end{equation}
The remaining smooth dependence on $s$ is contained in the subtraction polynomials $P_i(s)$. 

\subsection{$\boldsymbol{D}$-wave representation}
For the $D$-wave we work with a single-channel isoscalar $J=2$ $\pi\pi$ Omn\`es function.  It implements the right-hand cut generated by the $\pi\pi$ $D$-wave phase shift and contains the phase motion associated with the $f_2(1270)$ resonance. It is given by
\begin{equation}
	\Omega_2^0(s)=
	\exp\left\{\frac{s}{\pi}\int_{4M_{\pi}^2}^{\infty}
	\frac{\diff s'}{s'}\,\frac{\delta_2^0(s')}{s'-s}\right\},
\end{equation}
where $\delta_2^0$ is the isoscalar $\pi\pi$ $D$-wave phase shift.  In the numerical implementation, this input is taken from global fit 1 of Ref.~\cite{Pelaez:2025gpp}, which is the preferred dispersively constrained solution for the isoscalar $D$-wave, up to $\sqrt{s}=1.6\GeV$; above this energy it is smoothly continued to $\pi$ in the Omn\`es integral. A coupled-channel $D$-wave Omn\`es treatment can be used for simultaneous $\pi\pi/K\bar K$ analyses, especially when the $f_2'(1525)$ region is included~\cite{Danilkin:2025kyo}.  Since the present analysis focuses on the  $2^{++}$ multipoles at low energies up to $1.45\GeV$, we keep the $D$-wave final-state interaction in the pion channel.

After separating the threshold and soft-photon factors in Eq.~\eqref{eq:dwave-helicity-constraints}, the dispersive representation is applied to the basis amplitudes $h^0_{2,i}$. For the optional LHC test, we use the once-subtracted single-channel MO form
\begin{align}
	h^0_{2,i}(s)
	&=h^{0,L}_{2,i}(s)
	+\Omega_2^0(s)
	\notag\\
	&\quad\times
	\left(P_i^D(s)-\Delta^{L}_{2,i}(s;s_0)\right),
	\qquad i=1,2,3\, ,
\end{align}
where
\begin{align}
	\Delta^{L}_{2,i}(s;s_0)
	&=\frac{s-s_0}{\pi}\int_{4M_{\pi}^2}^{\infty}\diff s'
	\frac{\Im\left(\Omega_2^0(s')\right)^{-1}}{(s'-s_0)(s'-s)}
	h^{0,L}_{2,i}(s'),\notag\\
	h^{0,L}_{2,i}(s)
	&=h^{0,b_1}_{2,i}(s)\, .
\end{align}
The function $h^{0,b_1}_{2,i}$ is the combination of the $b_1$-exchange projections of App.~\ref{app:LHCs} that enter the basis amplitudes $h^0_{2,i}$.  The functions $\Delta^{L}_{2,i}$ denote the corresponding rescattering corrections.

In the fits, both $\boldsymbol{P}(s)$ and $P_i^{D}(s)$ are chosen as linear polynomials. The LHC rescattering terms $\boldsymbol{\Delta}^{L}_{0,++}$ and $\Delta^{L}_{2,i}$, however, are evaluated with one subtraction, which is sufficient for convergence of the spin-one-exchange contributions. An equivalent form with an additional subtraction would differ only by a polynomial in $s$ and can be absorbed into $\boldsymbol{P}(s)$ and $P_i^{D}(s)$. Since the explicit LHC terms are numerically small, they are not intended to describe the $S$- and $D$-wave intensities on their own, but rather to estimate their interference with the dominant production mechanism.

\section{Numerical results}
\label{sec:results}

In this section we compare the partial-wave dispersive representations with the BESIII mass-independent analysis of $J/\psi\to\gamma\pi^0\pi^0$~\cite{BESIII:2015rug}. The BESIII intensities are normalized with the total intensity yield and the determined branching fraction, which fixes the absolute scale of the fitted amplitudes.  The comparison is then organized in two steps.  We first fit the $0^{++}$ intensity, the three $2^{++}$ multipole intensities, and the total spectrum.  We then add the measured $0^{++}-2^{++}$ $E1$ phase difference to the same intensity data, so that the $S$- and $D$-wave amplitudes are tested also through their relative phase.  Comparing the two fits shows the phenomenological impact of the measured phase information, in particular its deviation from the elastic Watson-theorem expectation.

\subsection{Input data and fit strategy}

The input consists of the mass-independent  BESIII multipole intensities
\begin{equation}
	I^{0,E1}\, ,\qquad I^{2,X}\,\quad (X=E1,M2,E3)\, ,
\end{equation}
together with the total intensity $I^{\rm tot}$ and the measured $0^{++}-2^{++}$ $E1$ phase difference.  Here $I^{0,E1}$ denotes the $0^{++}$ $E1$ contribution, while $I^{2,X}$ denotes the contributions from the three $2^{++}$ radiative multipoles. The total intensity is related to the quoted component intensities by
\begin{equation}
	I_i^{\rm tot}=I_i^{0,E1}+\sum_{X}I_i^{2,X}\,, \qquad X=E1,M2,E3 \,.
	\label{eq:total-intensity-components}
\end{equation}
Here $i$ labels the experimental $\pi^0 \pi^0$ mass bin.

For these quantities, BESIII provides statistical uncertainties only in diagonal form.  In addition, BESIII quotes an overall systematic normalization uncertainty of $5.4\%$ for the results of the mass-independent analysis~\cite{BESIII:2015rug}.  The dominant contributions arise from the photon-detection efficiency ($5.0\%$) and the number of $J/\psi$ events ($0.8\%$), which we treat as normalization uncertainties that are fully correlated across all fitted intensity data points.  We assume the remaining systematic effects to be localized in individual energy bins and add them in quadrature to the diagonal covariance entries.  To avoid a potential bias associated with these multiplicative normalization uncertainties~\cite{DAgostini:1993arp}, we follow the iterative strategy of Refs.~\cite{Ball:2009qv,Hoferichter:2019gzf,Hoid:2020xjs}.  Starting from the empirical covariance matrix, we update it using the fitted intensity spectrum until the iteration converges to the unbiased minimum.
 
The BESIII mass-independent analysis is affected by mathematical ambiguities when extracting partial-wave amplitudes from intensities~\cite{Barrelet:1971pw}. Since the fitted intensity depends on quadratic combinations of the complex amplitudes, different amplitude coefficients can reproduce the same total intensity in a given mass bin.  In this regard, the trivial overall phase ambiguity is reduced by choosing the $2^{++}$ $E1$ amplitude to be real and positive. Furthermore, an arbitrary convention is chosen such that the $0^{++}-2^{++}$ $E1$ phase difference is constrained to be positive.  A nontrivial ambiguous partner can still remain when the analysis is truncated to the $0^{++}$ and $2^{++}$ amplitudes.  In the BESIII construction, this ambiguity is tied to the possibility that the
$2^{++}$ $E1$, $M2$, and $E3$ components have different phases. Below the $K\bar K$ threshold, BESIII assumes that the three $2^{++}$ multipoles share a common phase and the nontrivial ambiguous partner is therefore absent. Above the $K\bar K$ threshold, this assumption is relaxed, leading to the two distinct solutions observed by BESIII in several mass bins~\cite{BESIII:2015rug}. We therefore fit the two solutions to estimate which one is favored by dispersive constraints.  In contrast, the total intensity is determined directly from the event yield in each mass bin and does not depend on the choice of ambiguous solutions.  Its uncertainties are also much smaller than the diagonal errors provided for the different multipoles.  Therefore, it should be used as a common constraint on the fitted sum of components.

To put the binned BESIII intensities on an absolute scale, we follow the normalization procedure of Ref.~\cite{Danilkin:2025kyo} and use the  determined  branching fraction~\cite{BESIII:2015rug,ParticleDataGroup:2024cfk}
\begin{equation}
    \mathcal{B}(J/\psi\to\gamma\pi^{0}\pi^{0})=1.15(0)_\text{stat}(5)_\text{syst}\times10^{-3}\,,
\end{equation} 
where the statistical uncertainty is completely negligible, while the systematic uncertainty is an overall
normalization uncertainty  dominated by the photon detection efficiency.  Together with the bin width $\Delta M=15\MeV$, this gives
\begin{equation}
	\left(\frac{\diff\Gamma^{J,X}_{J/\psi\to\gamma\pi^0\pi^0}}{\Gamma_{J/\psi}\,\diff\sqrt{s}}\right)_i^{\rm data}
	=\frac{{\cal B}(J/\psi\to\gamma\pi^0\pi^0)}{\Delta M}\,\frac{I^{J,X}_i}{\sum_k I^{\rm tot}_k}\, ,
\end{equation}
where $\Gamma_{J/\psi}$ is the $J/\psi$ decay width, and the sum over \(k\) runs over the full BESIII mass range. Here $X=E1$ for $J=0$ and $X=E1,M2,E3$ for $J=2$.   The total spectrum is normalized by the same expression with $I_i^{J,X}$ replaced by $I_i^{\rm tot}$.  With this convention, the total spectrum integrates to the published branching fraction, while the multipole spectra are normalized to the same total yield. 

The differential branching fractions in the physical $\pi^0\pi^0$ channel are given by
\begin{align}
	\frac{\diff \Gamma^{0,E1}_{J/\psi\to\gamma\pi^0\pi^0}}{\Gamma_{J/\psi}\,\diff \sqrt{s}}
	&=\frac{1}{\Gamma_{J/\psi}}\,
	\frac{(q^2-s)\sqrt{s-4M_{\pi^0}^2}}
	{384\pi^3 (q^2)^{3/2}}
	\left|h^{n}_{0,E1}(s)\right|^2,\notag\\
	\frac{\diff \Gamma^{2,X}_{J/\psi\to\gamma\pi^0\pi^0}}{\Gamma_{J/\psi}\,\diff \sqrt{s}}
	&=\frac{1}{\Gamma_{J/\psi}}\,
	\frac{(q^2-s)\sqrt{s-4M_{\pi^0}^2}}
	{384\pi^3 (q^2)^{3/2}}
	5\left|\,h^{n}_{2,X}(s)\right|^2\,,\notag\\
    X&=E1,M2,E3\, .
\end{align}
Here $h^{n}_{0,E1}$ and $h^{n}_{2,X}$ denote the physical neutral-channel amplitudes defined in Sect.~\ref{sec:form}.  The total spectrum is not fitted as an independent amplitude, but as the sum
\begin{equation}
	\frac{\diff \Gamma^{\rm tot}_{J/\psi\to\gamma\pi^0\pi^0}}{\Gamma_{J/\psi}\,\diff \sqrt{s}}
	=
	\frac{\diff \Gamma^{0,E1}_{J/\psi\to\gamma\pi^0\pi^0}}{\Gamma_{J/\psi}\,\diff \sqrt{s}}
	+\sum_{X}
	\frac{\diff \Gamma^{2,X}_{J/\psi\to\gamma\pi^0\pi^0}}{\Gamma_{J/\psi}\,\diff \sqrt{s}} \,.
\end{equation}
Since the BESIII mass-independent analysis is performed in finite $M_{\pi^0\pi^0}$ bins, with the dynamical functions taken to be constant within each bin, the theory prediction is averaged over the corresponding bin interval,
\begin{equation}
	\frac{1}{\Delta M}\int_{\sqrt{s_i}-\Delta M /2}^{\sqrt{s_i}+\Delta M /2}\diff\sqrt{s}\,
	\frac{\diff\Gamma^{\alpha}_{J/\psi\to\gamma\pi^0\pi^0}}{\Gamma_{J/\psi}\,\diff\sqrt{s}}\,  .
\end{equation}
The finite-bin correction is implemented with a numerically faster iterative reweighting method following recent applications~\cite{Hoferichter:2025lcz,Hoid:2026atr}.

We consider two fit scenarios.  The first uses only the intensity information, while  the second adds the measured $0^{++}-2^{++}$ $E1$ phase difference,
\begin{align}\label{eq:chi2}
	\chi^2_{\rm int}
	&=\chi^2_{0,E1}+\chi^2_{2,E1}+\chi^2_{2,M2}
	+\chi^2_{2,E3}+\chi^2_{I^{\rm tot}}\, ,\notag\\
	\chi^2_{\rm int+ph}
	&=\chi^2_{\rm int}+\chi^2_{\Delta\phi}\, ,
\end{align}
where $\chi^2_{\Delta\phi}$ denotes the contribution from  the $0^{++}-2^{++}$ $E1$ phase difference.

%
\subsection{Combined intensity fits}

For the first fit scenario, the $S$-wave amplitude is described with the coupled-channel $\pi\pi/K\bar K$ MO representation of Sect.~\ref{sec:disp-reps}.  The subtraction polynomials are taken as real linear functions,
\begin{equation}
	P_i(s)=a_i+b_i s\, ,\qquad i=\pi,K\, ,
\end{equation}
which multiply the two columns of the Omn\`es matrix and allow production in both coupled channels before rescattering into the observed $\pi\pi$ channel.

For the $D$-wave amplitudes we use the single-channel  Omn\`es input $\Omega_2^0(s)$ detailed in Sect.~\ref{sec:disp-reps}.  The main fit is performed for the basis amplitudes $h^0_{2,i}$ of Eq.~\eqref{eq:dwave-helicity-constraints}, with subtraction polynomials
\begin{equation}
	P_i^D(s)=a_i^{D}+b_i^{D}s\,,\qquad i=1,2,3\, .
\end{equation}
In the fit, the helicity amplitudes are reconstructed with the kinematic relations in Eq.~\eqref{eq:dwave-helicity-constraints} and transformed to $E1$, $M2$, and $E3$ through Eq.~\eqref{eq:helicity-multipole}.  Thus, in this construction, the $D$-wave kinematic constraints are imposed before the comparison is made in the BESIII multipole basis.

The $b_1$ LHC term was tested to assess LHC interference with the dominant production terms. Its inclusion does not improve the fit quality and shifts the fitted parameters by amounts within their quoted uncertainties. We therefore do not include explicit LHC terms and set them to zero in the central results.

The $S$- and $D$-wave components are fitted simultaneously in the region where both representations are valid, $\sqrt{s}\leq 1.05\GeV$. In this common region, the theory prediction for the total spectrum is constructed from the same component sum as in Eq.~\eqref{eq:total-intensity-components}, so that the measured total spectrum provides a stringent constraint on the shared set of amplitudes.  This is particularly important in the transition from the $S$-wave-dominated low-energy region to the spectrum dominated by the $2^{++}$ multipoles.  Above $1.05\GeV$, the S-wave representation is not used; the $2^{++}$ multipole fit is continued up to $1.45\GeV$, where the single-channel $D$-wave representation remains adequate for the present purpose and the data are dominated by the $f_2(1270)$ structure.

\begin{figure}[!t]
	\centering
	\includegraphics[width=\linewidth]{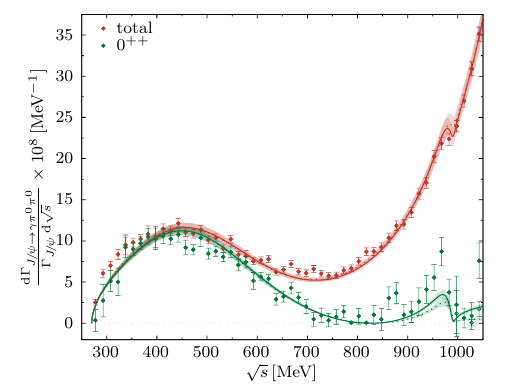}
	\caption{Combined intensity fit to the $\pi^0\pi^0$ input.  The total intensity is shown together with the $S$-wave contribution. The solid and dashed lines show fits to the first nominal solution and its ambiguous partner, respectively. The error bars on the data points show only statistical uncertainties.}
	\label{fig:combined-intensity-total}
\end{figure}

\begin{figure}[!t]
	\centering
	\includegraphics[width=\linewidth]{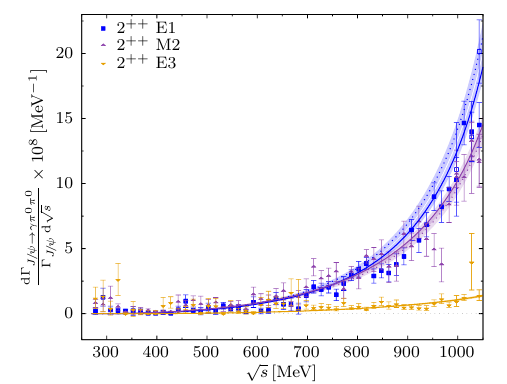}\\[0.4em]
	\includegraphics[width=\linewidth]{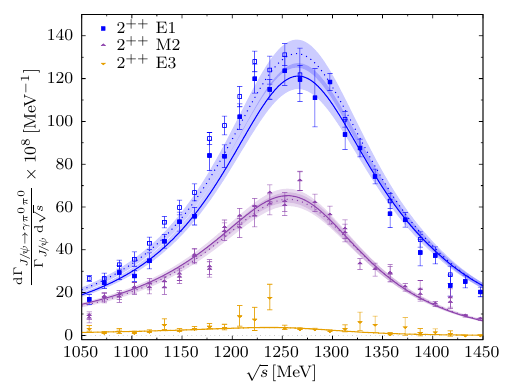}
	\caption{$2^{++}$ intensity contributions in the combined fit.  The upper panel shows the low-energy region, while the lower panel shows the high-energy region dominated by the $f_2(1270)$. The solid and dashed lines represent fits to the first nominal solution and its ambiguous partner.}
	\label{fig:combined-intensity-dwave}
\end{figure}

Figure~\ref{fig:combined-intensity-total} shows the fits to the total intensity, together with the corresponding $S$-wave contribution up to $1.05\GeV$.  The individual $2^{++}$ multipole intensities are shown in Fig.~\ref{fig:combined-intensity-dwave}, where the low- and high-energy regions are displayed separately to keep both the threshold behavior and the $f_2(1270)$ region visible. The fit parameters satisfying the $D$-wave kinematic constraints are summarized in Table~\ref{tab:total-kinematic-fit}.

\begin{table}[t]
	\renewcommand{\arraystretch}{1.25}
	\centering
	\begin{tabular}{l@{\hspace{2em}}r@{\hspace{2em}}r}
	\hline\hline
	 & nominal 1 & nominal 2 \\
		\hline
		\multirow[t]{2}{*}{$\chi^2/\text{dof}$}
			& $548.78/331$ & $642.11/331$ \\
			& $=1.66$ & $=1.94$ \\
				$p$-value & $5.15\times 10^{-13}$ & $3.84\times 10^{-22}$ \\
				$a_\pi\,[10^{-3}\GeV^{-2}]$ & $5.05(13)$ & $5.14(13)$ \\
				$b_\pi\,[10^{-3}\GeV^{-4}]$ & $-8.28(38)$ & $-6.61(37)$ \\
				$a_K\,[10^{-3}\GeV^{-2}]$ & $8.0(1.1)$ & $2.0(1.1)$ \\
				$b_K\,[10^{-3}\GeV^{-4}]$ & $-10.9(1.3)$ & $-3.7(1.3)$ \\
				$a^D_1\,[10^{-5}\GeV^{-4}]$ & $9.5(1.9)$ & $19.4(1.9)$ \\
				$b^D_1\,[10^{-5}\GeV^{-6}]$ & $4.8(1.1)$ & $-0.1(1.1)$ \\
				$a^D_2\,[10^{-5}\GeV^{-6}]$ & $5.44(24)$ & $4.44(23)$ \\
				$b^D_2\,[10^{-5}\GeV^{-8}]$ & $-2.32(14)$ & $-1.81(13)$ \\
				$a^D_3\,[10^{-5}\GeV^{-8}]$ & $-1.147(38)$ & $-1.050(36)$ \\
				$b^D_3\,[10^{-5}\GeV^{-10}]$ & $0.439(20)$ & $0.389(19)$ \\
	\hline\hline
	\end{tabular}
	\renewcommand{\arraystretch}{1.0}
	\caption{Combined intensity fits to the $\pi^0\pi^0$ input using the $h^0_{2,i}$ $D$-wave implementation, in which the kinematic constraints are imposed before transforming to the multipole basis.}
	\label{tab:total-kinematic-fit}
\end{table}

The values of $\chi^2/\text{dof}$ in Table~\ref{tab:total-kinematic-fit} are larger than those obtained from fits to the individual component spectra.  This is primarily because the combined fit also includes the total intensity, whose smaller statistical uncertainties impose a stringent constraint on the common sum of the $S$- and $D$-wave contributions.  Fits to the $0^{++}$ intensity and to the $2^{++}$ multipole intensities separately give $\chi^2/\text{dof}\simeq1.3$.   Including the systematic uncertainties does not remove this tension, since the bias-free iterative treatment of multiplicative normalization uncertainties largely affects the fit uncertainties and has little impact on the fitted spectral shape. The fit results show that the first nominal solution gives a lower $\chi^2$ than its ambiguous partner, with $\Delta\chi^2\simeq93$ for the same number of degrees of freedom.   Since only diagonal statistical uncertainties are available, the absolute $p$-values should not be overinterpreted; nevertheless, the relative comparison is meaningful within the same fit setup and indicates that the first solution is more compatible with the dispersive representation.  The leading $S$-wave polynomial, which is most relevant for the low-energy region,  is stable between the two solutions. Although the $0^{++}$ ambiguity occurs only in three bins in the analysis region, the second $S$-wave polynomial changes substantially in central value through $a_K$ and $b_K$.  Among the $2^{++}$ amplitudes, the largest changes occur in the first two $D$-wave polynomial blocks, including a sign change in $b_1^D$, consistent with the visible differences in the $E1$ and $M2$ intensities in Fig.~\ref{fig:combined-intensity-dwave}.  By contrast, the third $D$-wave block is much more stable.

To test the importance of the full $D$-wave kinematic constraints derived in Eq.~\eqref{eq:dwave-helicity-constraints}, we also perform a direct multipole fit with independent linear polynomials for the three $2^{++}$ multipoles,
\begin{align}
	h^0_{2,X}(s)
	&=(s-4M_\pi^2)(s-q^2)^{n_X}
	\Omega_2^0(s)\,\widetilde P_X^D(s),\notag\\
	\widetilde P_X^D(s)&=\widetilde a_X^D+\widetilde b_X^D s\, ,
\end{align}
with $X=E1,M2,E3$. The powers are chosen as $n_{E1}=1$, $n_{M2}=2$, and $n_{E3}=3$. This form builds in the expected leading zeros of the individual multipoles, but it treats the three $2^{++}$ multipoles as independent functions after these factors are removed.  It therefore does not fulfill the full set of $D$-wave kinematic constraints of Eq.~\eqref{eq:dwave-helicity-constraints}.  The resulting combined intensity fits are significantly worse, with $\chi^2/\text{dof}\approx2.9$ for the two nominal solutions.  The data therefore favor the fits in which the $D$-wave kinematic constraints are imposed, and we use the $h^0_{2,i}$ construction as the main result.

\subsection{Combined intensity and phase fits}

In addition to the intensities, the BESIII mass-independent analysis provides phase differences with respect to the $2^{++}$ $E1$ reference amplitude. The most relevant one for the present analysis is the $0^{++}-2^{++}$ $E1$ phase difference in the low-mass region.  An arbitrary convention is chosen such that the $0^{++}-2^{++}$ $E1$ phase difference is positive~\cite{BESIII:2015rug}.  However, owing to the ambiguities of the system, the corresponding negative solution is also a valid solution, as documented by BESIII.

\begin{figure}[!t]
	\centering
	\includegraphics[width=\linewidth]{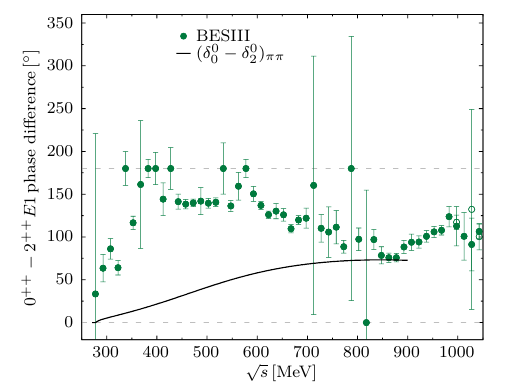}\\[0.4em]
	\includegraphics[width=\linewidth]{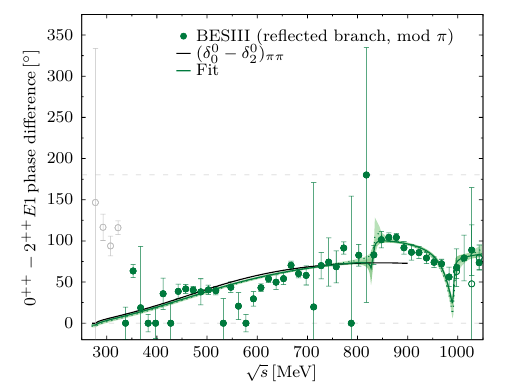}
	\caption{Measured $0^{++}-2^{++}$ $E1$ phase difference compared with the $\pi\pi$ phase-shift difference $\delta_0^0-\delta_2^0$ obtained from the Omn\`es inputs (black solid line).  The upper panel uses the experimental presentation convention.  The lower panel shows the reflected-sign branch mapped to the interval $[0,\pi)$. Mathematically, $\Delta \phi_{\mathrm{refl}} = \operatorname{mod}\left(-\Delta \phi_{\mathrm{BESIII}}, \pi\right) = \pi - \Delta \phi_{\mathrm{BESIII}}$, for $0 < \Delta \phi_{\mathrm{BESIII}} < \pi$.  In the lower panel, the solid and dashed green lines represent the combined intensity and phase fits to the first nominal solution and its ambiguous partner, respectively. The first four points above threshold are excluded from the fit.}
	\label{fig:phase-comparison}
\end{figure}

We first adopt the experimental presentation convention and compare the measured phase difference with the phase difference generated by the single-channel $S$- and $D$-wave Omn\`es inputs~\cite{Danilkin:2018qfn,Pelaez:2025gpp}.  Since the Omn\`es functions encode the $\pi\pi$ final-state interactions, this provides a direct comparison with the corresponding difference $\delta_0^0-\delta_2^0$ between the $\pi\pi$ scattering phase shifts. The result is shown in the upper panel of Fig.~\ref{fig:phase-comparison}.  Restricting the comparison to $\sqrt{s}\leq0.9\GeV$, where this elastic benchmark is applicable, we find an average deviation of $9.6\sigma$ from the central curve, similar to the observation of Ref.~\cite{Rodas:2022jpsi}.

We next investigate the negative solution allowed by the BESIII ambiguity, which has not been considered in previous analyses.  It is obtained by reversing the sign of the phase differences displayed in the experimental convention.  To compare this solution with the Omn\`es prediction, we recall that elastic unitarity determines the phase of a single-channel production amplitude only modulo $\pi$: multiplying the amplitude by an overall minus sign leaves the unitarity relation unchanged while shifting its phase by $\pi$~\cite{Oller:2019rej}.

We use this freedom to map the negative solution to the natural interval $[0,\pi)$ and compare it with the phase difference obtained from the Omn\`es inputs.  The result is shown in the lower panel of Fig.~\ref{fig:phase-comparison}.  This version is substantially more compatible with the elastic benchmark, with an average deviation of $3.3\sigma$ from the central curve, although some large fluctuations remain.

Since the negative solution agrees substantially better with the phase difference generated by the Omn\`es inputs, we use it in the combined fit and add $\chi^2_{\Delta\phi}$ to $\chi^2_{\rm int}$, as shown in Eq.~\eqref{eq:chi2}. The dispersive representations are kept the same as in the intensity fit, but the $S$-wave subtraction polynomials are promoted to complex functions,
\begin{equation}
	P_i(s)=e^{i\phi_{i}}\left(a_i+b_i s\right),\qquad i=\pi,K\,  .
\end{equation}
Here the coefficients $a_i$ and $b_i$ are real fit parameters, while the phases $\phi_i$ describe additional phases of the two $S$-wave subtraction components on top of the phase motion generated by the Omn\`es representation.  This fit strategy is aligned with the BESIII phase convention: the $2^{++}$ $E1$ amplitude is chosen to be real.  We therefore keep the $2^{++}$ polynomial functions real, so that their phase motion is generated only by the $J=2$ Omn\`es function.  We require the fitted phase difference to vary continuously with energy. Since systematic uncertainties cannot be assigned reliably to the measured phase difference, we use only statistical uncertainties for both the intensities and phases in this fit, so that the two data sets are treated on the same footing.

The first four near-threshold phase points are excluded from the phase fit because the partial-wave separation appears unstable in this region. Close to threshold, the $D$-wave multipoles are strongly suppressed by the kinematic factors in Eq.~\eqref{eq:dwave-helicity-constraints}, so the spectrum is expected to be dominated by the $S$-wave. The extracted $D$-wave intensities nevertheless fluctuate noticeably in the first bins, which makes the corresponding $0^{++}-2^{++}$ $E1$ phase difference unreliable for a smooth dispersive fit.

\begin{figure}[!t]
	\centering
	\includegraphics[width=\linewidth]{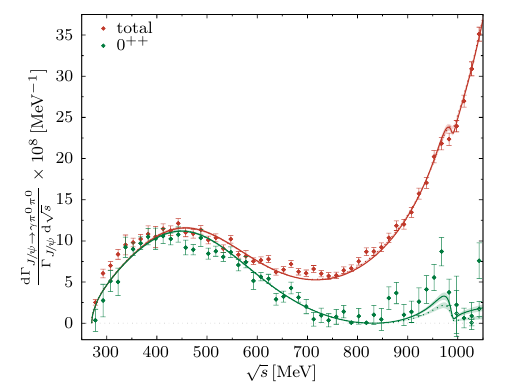}
	\caption{Total intensity in the combined intensity and phase fit, shown together with the corresponding $S$-wave contribution.}
	\label{fig:combined-phase-intensity-total}
\end{figure}

\begin{figure}[!t]
	\centering
	\includegraphics[width=\linewidth]{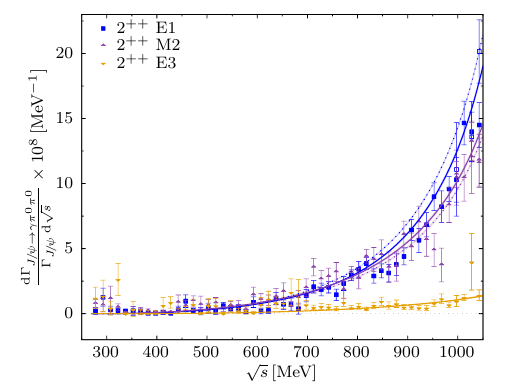}\\[0.4em]
	\includegraphics[width=\linewidth]{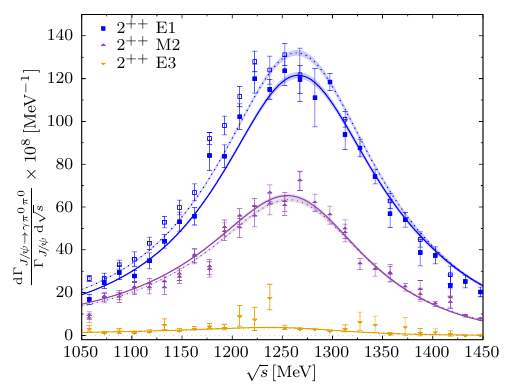}
	\caption{$2^{++}$ intensity contributions in the combined intensity and phase fit.  The upper panel shows the low-energy region, while the lower panel shows the high-energy region dominated by the $f_2(1270)$.}
	\label{fig:combined-phase-intensity-dwave}
\end{figure}

The fitted $0^{++}-2^{++}$ $E1$ phase difference is shown in the lower panel of Fig.~\ref{fig:phase-comparison}.  The corresponding fitted intensity spectra are displayed in Figs.~\ref{fig:combined-phase-intensity-total} and~\ref{fig:combined-phase-intensity-dwave}: the first compares the total intensity with the $S$-wave contribution, while the second shows the individual $2^{++}$ multipole intensities. The corresponding fit parameters are summarized in Table~\ref{tab:combined-phase-fit}.

\begin{table}[t]
	\renewcommand{\arraystretch}{1.25}
	\centering
	\begin{tabular}{l@{\hspace{2em}}r@{\hspace{2em}}r}
	\hline\hline
	 & nominal 1 & nominal 2 \\
	\hline
		\multirow[t]{2}{*}{$\chi^2/\text{dof}$}
			& $701.84/377$ & $810.70/377$ \\
			& $=1.86$ & $=2.15$ \\
			$p$-value & $7.24\times 10^{-22}$ & $7.98\times 10^{-34}$ \\
			$a_\pi\,[10^{-3}\GeV^{-2}]$ & $5.055(38)$ & $5.118(34)$ \\
			$b_\pi\,[10^{-3}\GeV^{-4}]$ & $-7.52(35)$ & $-6.68(10)$ \\
				$a_K\,[10^{-3}\GeV^{-2}]$ & $5.7(1.0)$ & $2.41(16)$ \\
				$b_K\,[10^{-3}\GeV^{-4}]$ & $-8.1(1.2)$ & $-4.03(45)$ \\
					$\phi_\pi\,[\text{rad}]$ & $-0.036(27)$ & $-0.060(24)$ \\
					$\phi_K\,[\text{rad}]$ & $-0.154(33)$ & $-0.134(28)$ \\
				$a^D_1\,[10^{-5}\GeV^{-4}]$ & $9.6(1.8)$ & $20.2(20)$ \\
				$b^D_1\,[10^{-5}\GeV^{-6}]$ & $4.8(1.1)$ & $-0.5(12)$ \\
				$a^D_2\,[10^{-5}\GeV^{-6}]$ & $5.43(22)$ & $4.35(25)$ \\
				$b^D_2\,[10^{-5}\GeV^{-8}]$ & $-2.32(13)$ & $-1.76(15)$ \\
				$a^D_3\,[10^{-5}\GeV^{-8}]$ & $-1.147(31)$ & $-1.042(34)$ \\
				$b^D_3\,[10^{-5}\GeV^{-10}]$ & $0.439(19)$ & $0.384(21)$ \\
	\hline\hline
	\end{tabular}
	\renewcommand{\arraystretch}{1.0}
	\caption{Combined intensity and phase fits to the $\pi^0\pi^0$ input using the same dispersive amplitudes as in the intensity-only fit.}
	\label{tab:combined-phase-fit}
\end{table}

The phase information provides an independent constraint beyond the intensity spectra.  The latter mainly constrain moduli and angle-integrated sums, while the measured phase difference probes the relative phase motion between the coupled-channel $S$-wave amplitude and the $2^{++}$ $E1$ amplitude.  A fit can therefore describe the $0^{++}$ and $2^{++}$ intensities separately, but still fail once the phase information is included. Once the appropriate phase solution is selected and the intensity constraints are included, the additional phases remain small and follow the same pattern for both nominal solutions, with negative values for both $\phi_\pi$ and $\phi_K$.  In particular, $\phi_K$ differs more clearly from zero, whereas $\phi_\pi$ remains a smaller correction. The rapid phase variation around $0.85\GeV$ is driven by the near-zero of the fitted $0^{++}$ amplitude required by the intensity data.

The first nominal solution remains favored after including the phase difference: the separation between the two solutions increases from $\Delta\chi^2=93$ in the intensity-only fits to $\Delta\chi^2=109$ in the combined intensity and phase fits. The leading $S$-wave coefficient $a_\pi$ is stable when the phase information is added.  For both nominal solutions, the largest changes occur in the kaon-channel $S$-wave polynomial parameters, with a smaller shift also visible in $b_\pi$.  This indicates that the measured phase difference is accommodated primarily through the kaon channel. The smaller parameter uncertainties in the combined intensity and phase fit should not be interpreted as arising from the additional phase constraint, since this fit uses only statistical uncertainties for the intensity data. 

We also considered a fit with additional phases $\phi_{b_i}$ that allow the constant and linear terms within each block to have different phases, but this implementation is disfavored by the data.

\subsection{Branching fractions}

Since the $f_0(500)$ is broad and the scalar structures overlap and interfere, a unique separation into individual resonance branching fractions is difficult. Following the window-integrated estimates of Refs.~\cite{Achasov:2020aun,Sakai:2019uig}, we therefore quote partial branching fractions obtained by integrating the corresponding regions of the spectrum.  The boundaries are chosen according to the structures visible in the BESIII spectrum and the fitted energy ranges.  We take the preferred fit to the first nominal solution as our central result and integrate the $S$-wave intensity over $2M_{\pi^0}<\sqrt{s}<0.9\GeV$ and $0.9\GeV<\sqrt{s}<1.05\GeV$ for the $f_0(500)$ and $f_0(980)$ regions, respectively.  For the $f_2(1270)$ region, we integrate the total $D$-wave intensity up to $\sqrt{s}=1.45\GeV$.  We obtain
\begin{align}
	\mathcal{B}\!\left(J/\psi\to\gamma f_0(500)\to\gamma\pi^0\pi^0\right)
	&=3.26(20)(8)\times10^{-5}\, ,\notag\\
	\mathcal{B}\!\left(J/\psi\to\gamma f_0(980)\to\gamma\pi^0\pi^0\right)
	&=0.27(4)(11)\times10^{-5}\, ,\notag\\
	\mathcal{B}\!\left(J/\psi\to\gamma f_2(1270)\to\gamma\pi^0\pi^0\right)
	&=4.58(29)(20)\times10^{-4}\, .
\end{align}	
Here the resonance labels identify the corresponding integration windows. The first uncertainty is propagated from the fit uncertainty and inflated by the PDG scale factor.  The second uncertainty is estimated by varying the coupled-channel Omn\`es input~\cite{Danilkin:2020pak}, the single-channel Omn\`es input~\cite{Pelaez:2025gpp}, the choice between the two nominal solutions, and the omission or inclusion of the phase information in the fit.  For each branching ratio, we assign the largest deviation from the central value among these variations as our systematic uncertainty. These values are consistent with the direct bin sums of Ref.~\cite{Achasov:2020aun}, which give $3.24\times10^{-5}$, $0.425\times10^{-5}$, and $4.82\times10^{-4}$, respectively, with a somewhat smaller $f_0(980)$ contribution in our fit driven by the inclusion of the total-intensity data.  Assuming isospin symmetry, the first two results correspond to $9.78(66)\times10^{-5}$ and $0.81(36)\times10^{-5}$ after summing over all $\pi\pi$ charge channels.  The $f_0(980)$ value is comparable to the result of Ref.~\cite{Sarantsev:2021ein}, while our window-integrated estimate for the $f_0(500)$ region is about one order of magnitude smaller than the value quoted there, $\mathcal{B}(J/\psi\to\gamma f_0(500)\to\gamma\pi\pi)=105(20)\times10^{-5}$.

As a separate estimate, we consider the kaon pole term followed by coupled-channel final-state interactions, as described in App.~\ref{app:LHCs}.  Using the unsubtracted representation applied in Ref.~\cite{Hoid:2026atr}, we find
\begin{align}
	\left.\mathcal{B}\!\left(J/\psi\to\gamma\pi^0\pi^0\right)\right|_{K\text{-pole}}
	&=2.58(39)\times10^{-7}\, ,\notag\\
	\left.\mathcal{B}\!\left(J/\psi\to\gamma f_0(980)\to\gamma\pi^0\pi^0\right)\right|_{K\text{-pole}}
	&=1.00(23)\times10^{-7}\, .
\end{align}
The uncertainties of these predictions are dominated by the coupled-channel Omn\`es input~\cite{Danilkin:2020pak}. Both values are one to two orders of magnitude smaller than the corresponding strengths in the fitted spectrum, confirming that the kaon-pole contribution is subleading. The result in the $f_0(980)$ region is consistent with the range $(0.26\text{--}1.04)\times10^{-7}$ obtained in Ref.~\cite{Xiao:2019lrj}.  The vector-meson-dominance mechanism considered in Ref.~\cite{Sakai:2019uig} gives the smaller neutral-channel estimate $1.2\times10^{-8}$. These estimates indicate that such mechanisms provide only subleading contributions and cannot account for the observed spectrum.

\section{Conclusions and outlook}
\label{sec:conc}
We have presented a dispersive analysis of the low-energy $\pi^0\pi^0$ system in $J/\psi\to\gamma\pi^0\pi^0$ using the intensities and relative phase information from the BESIII mass-independent amplitude analysis~\cite{BESIII:2015rug}.  We described the isoscalar $S$-wave with a coupled-channel $\pi\pi/K\bar K$ Muskhelishvili--Omn\`es  representation, which implements the strong final-state interactions in the $f_0(500)$ and $f_0(980)$ region.  The $D$-wave was treated with a single-channel $\pi\pi$ Omn\`es function, with the kinematic constraints imposed at the level of the helicity amplitudes before transforming to the experimental $2^{++}$ $E1$, $M2$, and $E3$ multipoles.  The analysis combines the total intensity, the $0^{++}$ intensity, all three measured $2^{++}$ multipoles, and the $0^{++}-2^{++}$ $E1$ phase difference in one dispersive framework.  We normalized the BESIII intensities using the total event yield and  the corresponding branching fraction, thereby fixing the absolute scale of the fitted amplitudes.

The comparison with the BESIII intensities shows the importance of including the total spectrum together with the individual $0^{++}$ and $2^{++}$ components.   It is free from mathematical ambiguities and has much smaller uncertainties than the individual multipole intensities. It therefore provides a strong constraint on the common set of  $0^{++}$ and $2^{++}$ components, and stabilizes the fits for the two nominal solutions.  Within the same fit setup, the first BESIII nominal solution is favored over its ambiguous partner, with a substantially lower $\chi^2$, although this comparison should be interpreted with the available diagonal statistical uncertainties.  The data also favor the construction in which the $D$-wave kinematic constraints are fully implemented; direct fits of independent $2^{++}$ multipoles with only the leading threshold and soft-photon zeros give significantly worse descriptions.  In addition, we provided a controlled estimate of the crossed-channel effects relevant in the low-energy region, finding that the left-hand-cut contributions are numerically negligible.

We further included the measured $0^{++}-2^{++}$ $E1$ phase difference in our analysis.  This observable gives information beyond the intensity spectra, because it probes the relative phase motion between the $S$-wave and the $2^{++}$ $E1$ amplitudes.  A key outcome is the identification of the negative solution of the BESIII phase ambiguity as the one compatible with elastic unitarity and the Omn\`es phase motion, once the modulo-$\pi$ freedom of production amplitudes is taken into account.  In the combined intensity and phase fit, this solution requires only small additional phases in the $S$-wave subtraction polynomials.  The measured phase information can therefore be reconciled with dispersive final-state interactions without introducing large additional phases.  This result supports a dispersive treatment in which analyticity, unitarity, and kinematic constraints are imposed before comparison with the binned mass-independent BESIII data.

For higher-mass extensions, the same kinematic constraints can be combined with a simultaneous analysis of the $\pi\pi$ and $K\bar K$ final states~\cite{BESIII:2015rug,BESIII:2018ubj}. Such an extension would test all $2^{++}$ multipoles in a common coupled-channel framework, going beyond the leading-$E1$ treatment of Ref.~\cite{Danilkin:2025kyo}, especially in the region where the $f_2'(1525)$ becomes relevant.  Experimentally, covariance information beyond the diagonal uncertainties, together with a systematic uncertainty estimate for the measured phase differences, would make the statistical interpretation of the dispersive constraints more conclusive.  The normalized amplitudes obtained here provide useful hadronic input beyond light-meson spectroscopy.  The same $0^{++}$ and $2^{++}$ $\pi\pi$ partial waves enter $t$-channel dispersive representations of two-pion contributions to gravitational form factors.  In particular, applications to near-threshold $J/\psi$ photoproduction and to the gluonic gravitational form factors of the proton require reliable time-like $\pi\pi$ amplitudes with controlled final-state interactions.  The present analysis is a step toward such applications, connecting radiative charmonium decay data to future dispersive studies of gravitational form factors.

\acknowledgments
We thank Jake Bennett for helpful correspondence. Financial support by the DFG through the fund provided to the Research Unit  ``Photon-photon interactions in the Standard Model and beyond'' (Projektnummer 458854507 - FOR 5327) is gratefully acknowledged.

\appendix
\section{Left-hand-cut contributions}
\label{app:LHCs}
The expressions below are provided for completeness. They were used only to estimate the size of possible LHC effects. Since all resulting contributions are numerically small, and even the largest $b_1$-exchange term does not affect the final fits, no explicit LHC term is included in the central results.

The leading spin-one exchanges considered here are the $\rho$ and $b_1$ exchanges, written in terms of vector and axial-vector vertices.  Here $P$ denotes a pseudoscalar field. For the vector case, relevant for $\rho$ exchange, we use
\begin{align}
	\mathcal{L}_{VP\gamma}&=e\,C_{VP\gamma}\,\epsilon^{\mu\nu\alpha\beta}
	F_{\mu\nu}\,\partial_\alpha P\,V_\beta\, ,\notag\\
	\mathcal{L}_{\psi VP}&=C_{\psi VP}\,\epsilon^{\mu\nu\alpha\beta}
	\partial_\mu\psi_\nu\,\partial_\alpha P\,V_\beta\, .
\end{align}
For the $b_1$, denoted by $B_\mu$, we use the $BP\gamma$ convention of Ref.~\cite{Garcia-Martin:2010kyn} and choose the analogous field-strength form for the $\psi BP$ coupling,
\begin{align}
	\mathcal{L}_{BP\gamma}&=e\,C_{BP\gamma}\,
	F_{\mu\nu}\,\partial^\mu B^\nu\,P\, ,\notag\\
	\mathcal{L}_{\psi BP}&=C_{\psi BP}\,
	\psi_{\mu\nu}\,\partial^\mu B^\nu\,P,\qquad
	\psi_{\mu\nu}=\partial_\mu\psi_\nu-\partial_\nu\psi_\mu\, .
\end{align}
The magnitudes of the radiative couplings are fixed from the corresponding decay widths.  With the above convention,
\begin{align}
	\Gamma_{V\to P\gamma}&=
	\frac{\alpha}{6}\,\left|C_{VP\gamma}\right|^2
	\frac{\left(M_V^2-M_P^2\right)^3}{M_V^3} ,\notag\\
	\Gamma_{B\to P\gamma}&=
	\frac{\alpha}{24}\,\left|C_{BP\gamma}\right|^2
	\frac{\left(M_B^2-M_P^2\right)^3}{M_B^3} .
\end{align}
The hadronic couplings are obtained analogously from the two-body widths
\begin{align}
	\Gamma_{J/\psi\to VP}&=
	\frac{\left|C_{\psi VP}\right|^2}{96\pi (q^2)^{3/2}}\,
	\lambda^{3/2}\!\left(q^2,M_V^2,M_P^2\right) ,\notag\\
	\Gamma_{J/\psi\to BP}&=
	\frac{\left|C_{\psi BP}\right|^2}{96\pi (q^2)^{3/2}}\,
	\lambda^{1/2}\!\left(q^2,M_B^2,M_P^2\right)
	\notag\\
	&\quad\times
	\left[\left(q^2+M_B^2-M_P^2\right)^2+2q^2M_B^2\right] ,
\end{align}
where $\lambda(x,y,z)=x^2+y^2+z^2-2(xy+xz+yz)$. Using the partial widths collected in Ref.~\cite{ParticleDataGroup:2024cfk}, we obtain
\begin{align}
	\left|C_{\rho\pi\gamma}\right|&=0.366(31)\,\GeV^{-1},\notag\\
	\left|C_{b_1\pi\gamma}\right|&=0.646(83)\,\GeV^{-1},\notag\\
	\left|C_{\psi\rho\pi}\right|&=2.67(13)\times10^{-3}\,\GeV^{-1},\notag\\
	\left|C_{\psi b_1\pi}\right|&=1.25(16)\times10^{-3}\,\GeV^{-1}.
\end{align}

For the vector exchange, $V=\rho$, we define
\begin{align}
	\sigma_P(s)&=\sqrt{1-\frac{4M_P^2}{s}}\, ,\notag\\
	t_\pm(s)&=M_P^2+\frac{q^2-s}{2}\left(1\pm\sigma_P(s)\right) ,
\end{align}
and define the logarithmic function
\begin{equation}
	L_V(s)=\log\frac{M_V^2-t_+(s)}{M_V^2-t_-(s)}\, .
\end{equation}
The pole part of the vector-exchange $S$-wave projection is
\begin{align}
	\mathcal{S}_V(s)
	&=
	\frac{L_V(s)}{\sigma_P(s)}
	\left[-2M_V^2+2q^2\left(\frac{M_V^2-M_P^2}{s-q^2}\right)^2\right]
	\notag\\
	&\quad
	+q^2\left[1-2\frac{M_V^2-M_P^2}{s-q^2}\right] .
\end{align}
The full $S$-wave projection contains in addition $\mathcal{S}_V^{\rm full}(s)=\mathcal{S}_V(s)+2(s-q^2)$.  Since this extra term is polynomial in $s-q^2$, it is absorbed into the production polynomial.  The $S$-wave amplitude is then
\begin{equation}
	h_{0,++}^{V}(s)
	=\frac{e\,C_{VP\gamma}C_{\psi VP}}{2}\,\mathcal{S}_V(s)\, .
\end{equation}
For the $D$-wave, we first define the functions entering the $J=2$ projection of $H_{++}$,
\begin{align}
	X_V(s)&=\frac{s-q^2+2M_V^2-2M_P^2}
	{(s-q^2)\sigma_P(s)}\, ,\notag\\
	F_V(s)&=M_V^2(s-q^2)^2-q^2\left(M_V^2-M_P^2\right)^2 .
\end{align}
The three $D$-wave projections are 
\begin{widetext}
\begin{align}
	\mathcal{D}_{++}^{V}(s)&=
	\frac{F_V(s)}{\sigma_P(s)(s-q^2)^2}
	\left[\left(1-3X_V^2(s)\right)L_V(s)+6X_V(s)\right] ,\notag\\
	\mathcal{D}_{+-}^{V}(s)&=
	\frac{\sqrt{6}}{8}\,s\,\sigma_P(s)
	\left[\left(1-X_V^2(s)\right)^2L_V(s)
	+\frac{10}{3}X_V(s)-2X_V^3(s)\right] ,\notag\\
	\mathcal{D}_{+0}^{V}(s)&=
	2\sqrt{3}\,M_{J/\psi}\sqrt{s}\,
	\frac{M_V^2-M_P^2}{s-q^2}
	\left[\frac{3X_V^2(s)-2}{3}
	-\frac{X_V(s)\left(X_V^2(s)-1\right)}{2}L_V(s)\right] .
\end{align}
\end{widetext}
The $D$-wave left-hand-cut pieces are
\begin{align}
	h_{2,++}^{V}(s)&=
	\frac{e}{2}\,C_{VP\gamma}C_{\psi VP}\,\mathcal{D}_{++}^{V}(s)\, ,\notag\\
	h_{2,+-}^{V}(s)&=
	\frac{e}{2}\,C_{VP\gamma}C_{\psi VP}\,\mathcal{D}_{+-}^{V}(s)\, ,\notag\\
	h_{2,+0}^{V}(s)&=
	\frac{e}{2}\,C_{VP\gamma}C_{\psi VP}\,\mathcal{D}_{+0}^{V}(s)\, .
\end{align}

For the axial-vector exchange, $B=b_1$, the same definitions of $\sigma_P(s)$ and $t_\pm(s)$ are used.  The corresponding logarithmic function is
\begin{equation}
	L_B(s)=\log\frac{M_B^2-t_+(s)}{M_B^2-t_-(s)}\, .
\end{equation}
The $S$-wave projection gives $\mathcal{S}_B^{\rm full}(s)=\mathcal{S}_B(s)-2(s-q^2)$, where the pole part retained in the left-hand cut is
\begin{align}
	\mathcal{S}_B(s)
	&=
	\frac{L_B(s)}{\sigma_P(s)}
	\left[2M_B^2+2q^2\left(\frac{M_B^2-M_P^2}{s-q^2}\right)^2\right]
	\notag\\
	&\quad
	+q^2\left[1-2\frac{M_B^2-M_P^2}{s-q^2}\right] .
\end{align}
The axial $S$-wave contribution is therefore
\begin{equation}
	h_{0,++}^{B}(s)
	=\frac{e}{4}\,C_{BP\gamma}C_{\psi BP}\,\mathcal{S}_B(s)\, .
\end{equation}
For the $D$-wave, the numerator entering the $J=2$ projection of $H_{++}$ differs from the vector case:
\begin{align}
	X_B(s)&=\frac{s-q^2+2M_B^2-2M_P^2}
	{(s-q^2)\sigma_P(s)}\, ,\notag\\
	F_B^{(+)}(s)&=M_B^2(s-q^2)^2+q^2\left(M_B^2-M_P^2\right)^2 .
\end{align}
Accordingly,
\begin{align}
	\mathcal{D}_{++}^{B}(s)
	&=
	-\frac{F_B^{(+)}(s)}{\sigma_P(s)(s-q^2)^2}
	\notag\\
	&\quad\times
	\left[\left(1-3X_B^2(s)\right)L_B(s)+6X_B(s)\right] .
\end{align}
The remaining $D$-wave projection functions keep the vector kinematic form after $M_V\to M_B$. The axial $D$-wave left-hand-cut pieces are
\begin{align}
	h_{2,++}^{B}(s)&=
	\frac{e}{4}\,C_{BP\gamma}C_{\psi BP}\,\mathcal{D}_{++}^{B}(s)\, ,\notag\\
	h_{2,+-}^{B}(s)&=
	\frac{e}{4}\,C_{BP\gamma}C_{\psi BP}\,\mathcal{D}_{+-}^{B}(s)\, ,\notag\\
	h_{2,+0}^{B}(s)&=
	\frac{e}{4}\,C_{BP\gamma}C_{\psi BP}\,\mathcal{D}_{+0}^{B}(s)\, .
\end{align}
The multipole amplitudes are then obtained from the helicity basis through Eq.~\eqref{eq:helicity-multipole}. For the physical exchanges one sets $V=\rho$ and $B=b_1$.  For these numerically subleading LHC estimates, the numerical implementation uses the phenomenological complex-mass replacement
\begin{equation}
	M_\rho\to M_\rho-\frac{i}{2}\Gamma_\rho\, ,\qquad
	M_{b_1}\to M_{b_1}-\frac{i}{2}\Gamma_{b_1}\, .
\end{equation}
rather than a dispersive finite-width treatment by a spectral representation~\cite{Hoid:2026atr,Zanke:2021wiq,Messerli:2025rnv}.

In addition, we estimate the charged-pseudoscalar Born terms.  They follow from scalar QED and the effective coupling
\begin{align}
	\mathcal{L}_{\psi P^+P^-}
	&=i g_{\psi P^+P^-}\,\psi_\mu
	\left[P^+\partial^\mu P^- - P^-\partial^\mu P^+\right]
	\notag\\
	&\quad
	+2e\,g_{\psi P^+P^-}\,\psi_\mu A^\mu P^+P^- ,
\end{align}
with $P=\pi,K$.
The coupling is fixed from
\begin{equation}
	\Gamma_{J/\psi\to P^+P^-}
	=\frac{\left|g_{\psi P^+P^-}\right|^2}{48\pi}M_{J/\psi}
	\left(1-\frac{4M_{P^+}^2}{q^2}\right)^{3/2}.
\end{equation}
Using the partial widths collected in Ref.~\cite{ParticleDataGroup:2024cfk}, this gives
\begin{align}
		\left|g_{\psi K^+K^-}\right|&=1.273(16)\times10^{-3}\, ,\notag\\
		\left|g_{\psi\pi^+\pi^-}\right|&=8.19(40)\times10^{-4}\, .
\end{align}
Following the notation of Ref.~\cite{Hoid:2026atr}, we define
\begin{align}
	\sigma_{P^+}(s)&=\sqrt{1-\frac{4M_{P^+}^2}{s}}\, ,\notag\\
	L_P(s)&=\frac{1}{\sigma_{P^+}(s)}
	\log\frac{1+\sigma_{P^+}(s)}{1-\sigma_{P^+}(s)}\, .
\end{align}
The charged-channel $S$-wave Born projections are then
\begin{align}
	k_{0,++}^{c,\text{Born}}(s)
	&=e\,g_{\psi K^+K^-}\,
	\frac{4M_{K^+}^2 L_K(s)-2q^2}{s-q^2}\, ,\notag\\
	h_{0,++}^{c,\text{Born}}(s)
	&=e\,g_{\psi\pi^+\pi^-}\,
	\frac{4M_{\pi^+}^2 L_\pi(s)-2q^2}{s-q^2}\, .
\end{align}
The charged-kaon $D$-wave Born projections are
\begin{widetext}
\begin{align}
	k_{2,++}^{c,\text{Born}}(s)
	&=-\frac{2e\,g_{\psi K^+K^-}M_{K^+}^2}{s-q^2}
	\left[
	\left(1-\frac{3}{\sigma_{K^+}^2(s)}\right)L_K(s)
	+\frac{6}{\sigma_{K^+}^2(s)}
	\right] ,\notag\\
	k_{2,+-}^{c,\text{Born}}(s)
	&=\frac{\sqrt{6}\,e\,g_{\psi K^+K^-}\,s}{4(s-q^2)}
	\left[
	\left(1-\frac{1}{\sigma_{K^+}^2(s)}\right)^2
	\sigma_{K^+}^2(s)L_K(s)
	-\frac{2}{\sigma_{K^+}^2(s)}+\frac{10}{3}
	\right] ,\notag\\
	k_{2,+0}^{c,\text{Born}}(s)
	&=\frac{2e\,g_{\psi K^+K^-}M_{J/\psi}\sqrt{s}}{\sqrt{3}(s-q^2)}
	\left[
	-2+\frac{3}{\sigma_{K^+}^2(s)}
	+\frac{3}{2}\left(1-\frac{1}{\sigma_{K^+}^2(s)}\right)L_K(s)
	\right] .
\end{align}
\end{widetext}
The charged-pion projections $h_{2,\lambda_1\lambda_2}^{c,\text{Born}}$ are obtained from  the charged-kaon expressions by the replacements
$g_{\psi K^+K^-}\to g_{\psi\pi^+\pi^-}$, $M_{K^+}\to M_{\pi^+}$, $\sigma_{K^+}\to\sigma_{\pi^+}$, and $L_K\to L_\pi$.

\bibliographystyle{apsrev4-1_mod_2}
\bibliography{ref}

\begin{thebibliography}{62}%
\makeatletter
\providecommand \@ifxundefined [1]{%
 \@ifx{#1\undefined}
}%
\providecommand \@ifnum [1]{%
 \ifnum #1\expandafter \@firstoftwo
 \else \expandafter \@secondoftwo
 \fi
}%
\providecommand \@ifx [1]{%
 \ifx #1\expandafter \@firstoftwo
 \else \expandafter \@secondoftwo
 \fi
}%
\providecommand \natexlab [1]{#1}%
\providecommand \enquote  [1]{``#1''}%
\providecommand \bibnamefont  [1]{#1}%
\providecommand \bibfnamefont [1]{#1}%
\providecommand \citenamefont [1]{#1}%
\providecommand \href@noop [0]{\@secondoftwo}%
\providecommand \href [0]{\begingroup \@sanitize@url \@href}%
\providecommand \@href[1]{\@@startlink{#1}\@@href}%
\providecommand \@@href[1]{\endgroup#1\@@endlink}%
\providecommand \@sanitize@url [0]{\catcode `\\12\catcode `\$12\catcode
  `\&12\catcode `\#12\catcode `\^12\catcode `\_12\catcode `\%12\relax}%
\providecommand \@@startlink[1]{}%
\providecommand \@@endlink[0]{}%
\providecommand \url  [0]{\begingroup\@sanitize@url \@url }%
\providecommand \@url [1]{\endgroup\@href {#1}{\urlprefix }}%
\providecommand \urlprefix  [0]{URL }%
\providecommand \Eprint [0]{\href }%
\providecommand \doibase [0]{http://dx.doi.org/}%
\providecommand \selectlanguage [0]{\@gobble}%
\providecommand \bibinfo  [0]{\@secondoftwo}%
\providecommand \bibfield  [0]{\@secondoftwo}%
\providecommand \translation [1]{[#1]}%
\providecommand \BibitemOpen [0]{}%
\providecommand \bibitemStop [0]{}%
\providecommand \bibitemNoStop [0]{.\EOS\space}%
\providecommand \EOS [0]{\spacefactor3000\relax}%
\providecommand \BibitemShut  [1]{\csname bibitem#1\endcsname}%
\let\auto@bib@innerbib\@empty
\bibitem [{\citenamefont {Ablikim}\ \emph {et~al.}(2015)\citenamefont {Ablikim}
  \emph {et~al.}}]{BESIII:2015rug}%
  \BibitemOpen
  \bibfield  {author} {\bibinfo {author} {\bibfnamefont {M.}~\bibnamefont
  {Ablikim}}  \emph {et~al.} (\bibinfo {collaboration} {BESIII}),\ }\href
  {\doibase 10.1103/PhysRevD.92.052003} {\bibfield  {journal} {\bibinfo
  {journal} {Phys. Rev. D}\ }\textbf {\bibinfo {volume} {92}},\ \bibinfo
  {pages} {052003} (\bibinfo {year} {2015})},\ \bibinfo {note} {[Erratum: Phys.
  Rev. D {\bf 93}, 039906 (2016)]},\ \Eprint {http://arxiv.org/abs/1506.00546}
  {arXiv:1506.00546 [hep-ex]}\BibitemShut {NoStop}%
\bibitem [{\citenamefont {Ablikim}\ \emph {et~al.}(2018)\citenamefont {Ablikim}
  \emph {et~al.}}]{BESIII:2018ubj}%
  \BibitemOpen
  \bibfield  {author} {\bibinfo {author} {\bibfnamefont {M.}~\bibnamefont
  {Ablikim}}  \emph {et~al.} (\bibinfo {collaboration} {BESIII}),\ }\href
  {\doibase 10.1103/PhysRevD.98.072003} {\bibfield  {journal} {\bibinfo
  {journal} {Phys. Rev. D}\ }\textbf {\bibinfo {volume} {98}},\ \bibinfo
  {pages} {072003} (\bibinfo {year} {2018})},\ \Eprint
  {http://arxiv.org/abs/1808.06946} {arXiv:1808.06946 [hep-ex]}\BibitemShut
  {NoStop}%
\bibitem [{\citenamefont {K{\"o}pke}\ and\ \citenamefont
  {Wermes}(1989)}]{Kopke:1988cs}%
  \BibitemOpen
  \bibfield  {author} {\bibinfo {author} {\bibfnamefont {L.}~\bibnamefont
  {K{\"o}pke}} and \bibinfo {author} {\bibfnamefont {N.}~\bibnamefont
  {Wermes}},\ }\href {\doibase 10.1016/0370-1573(89)90074-4} {\bibfield
  {journal} {\bibinfo  {journal} {Phys. Rept.}\ }\textbf {\bibinfo {volume}
  {174}},\ \bibinfo {pages} {67} (\bibinfo {year} {1989})}\BibitemShut
  {NoStop}%
\bibitem [{\citenamefont {Klempt}\ and\ \citenamefont
  {Zaitsev}(2007)}]{Klempt:2007cp}%
  \BibitemOpen
  \bibfield  {author} {\bibinfo {author} {\bibfnamefont {E.}~\bibnamefont
  {Klempt}} and \bibinfo {author} {\bibfnamefont {A.}~\bibnamefont {Zaitsev}},\
  }\href {\doibase 10.1016/j.physrep.2007.07.006} {\bibfield  {journal}
  {\bibinfo  {journal} {Phys. Rept.}\ }\textbf {\bibinfo {volume} {454}},\
  \bibinfo {pages} {1} (\bibinfo {year} {2007})},\ \Eprint
  {http://arxiv.org/abs/0708.4016} {arXiv:0708.4016 [hep-ph]}\BibitemShut
  {NoStop}%
\bibitem [{\citenamefont {Okubo}(1963)}]{Okubo:1963fa}%
  \BibitemOpen
  \bibfield  {author} {\bibinfo {author} {\bibfnamefont {S.}~\bibnamefont
  {Okubo}},\ }\href {\doibase 10.1016/S0375-9601(63)92548-9} {\bibfield
  {journal} {\bibinfo  {journal} {Phys. Lett.}\ }\textbf {\bibinfo {volume}
  {5}},\ \bibinfo {pages} {165} (\bibinfo {year} {1963})}\BibitemShut {NoStop}%
\bibitem [{\citenamefont {Zweig}(1964)}]{Zweig:1964jf}%
  \BibitemOpen
  \bibfield  {author} {\bibinfo {author} {\bibfnamefont {G.}~\bibnamefont
  {Zweig}},\ }in\ \href {\doibase 10.17181/CERN-TH-412} {\emph {\bibinfo
  {booktitle} {{DEVELOPMENTS IN THE QUARK THEORY OF HADRONS. VOL. 1. 1964 -
  1978}}}},\ \bibinfo {editor} {edited by\ \bibinfo {editor} {\bibfnamefont
  {D.~B.}\ \bibnamefont {Lichtenberg}} and \bibinfo {editor} {\bibfnamefont
  {S.~P.}\ \bibnamefont {Rosen}}}\ (\bibinfo {year} {1964})\ pp.\ \bibinfo
  {pages} {22--101}\BibitemShut {NoStop}%
\bibitem [{\citenamefont {Iizuka}(1966)}]{Iizuka:1966fk}%
  \BibitemOpen
  \bibfield  {author} {\bibinfo {author} {\bibfnamefont {J.}~\bibnamefont
  {Iizuka}},\ }\href {\doibase 10.1143/PTPS.37.21} {\bibfield  {journal}
  {\bibinfo  {journal} {Prog. Theor. Phys. Suppl.}\ }\textbf {\bibinfo {volume}
  {37}},\ \bibinfo {pages} {21} (\bibinfo {year} {1966})}\BibitemShut {NoStop}%
\bibitem [{\citenamefont {Bali}\ \emph {et~al.}(1993)\citenamefont {Bali},
  \citenamefont {Schilling}, \citenamefont {Hulsebos}, \citenamefont {Irving},
  \citenamefont {Michael},\ and\ \citenamefont {Stephenson}}]{Bali:1993fb}%
  \BibitemOpen
  \bibfield  {author} {\bibinfo {author} {\bibfnamefont {G.~S.}\ \bibnamefont
  {Bali}}, \bibinfo {author} {\bibfnamefont {K.}~\bibnamefont {Schilling}},
  \bibinfo {author} {\bibfnamefont {A.}~\bibnamefont {Hulsebos}}, \bibinfo
  {author} {\bibfnamefont {A.~C.}\ \bibnamefont {Irving}}, \bibinfo {author}
  {\bibfnamefont {C.}~\bibnamefont {Michael}},  \emph {et~al.} (\bibinfo
  {collaboration} {UKQCD}),\ }\href {\doibase 10.1016/0370-2693(93)90948-H}
  {\bibfield  {journal} {\bibinfo  {journal} {Phys. Lett. B}\ }\textbf
  {\bibinfo {volume} {309}},\ \bibinfo {pages} {378} (\bibinfo {year}
  {1993})},\ \Eprint {http://arxiv.org/abs/hep-lat/9304012}
  {arXiv:hep-lat/9304012}\BibitemShut {NoStop}%
\bibitem [{\citenamefont {Morningstar}\ and\ \citenamefont
  {Peardon}(1999)}]{Morningstar:1999rf}%
  \BibitemOpen
  \bibfield  {author} {\bibinfo {author} {\bibfnamefont {C.~J.}\ \bibnamefont
  {Morningstar}} and \bibinfo {author} {\bibfnamefont {M.}~\bibnamefont
  {Peardon}},\ }\href {\doibase 10.1103/PhysRevD.60.034509} {\bibfield
  {journal} {\bibinfo  {journal} {Phys. Rev. D}\ }\textbf {\bibinfo {volume}
  {60}},\ \bibinfo {pages} {034509} (\bibinfo {year} {1999})},\ \Eprint
  {http://arxiv.org/abs/hep-lat/9901004} {arXiv:hep-lat/9901004
  [hep-lat]}\BibitemShut {NoStop}%
\bibitem [{\citenamefont {Chen}\ \emph {et~al.}(2006)\citenamefont {Chen},
  \citenamefont {Alexandru}, \citenamefont {Dong}, \citenamefont {Draper},
  \citenamefont {Horvath}, \citenamefont {Lee}, \citenamefont {Liu},
  \citenamefont {Mathur}, \citenamefont {Morningstar}, \citenamefont {Peardon},
  \citenamefont {Tamhankar}, \citenamefont {Young},\ and\ \citenamefont
  {Zhang}}]{Chen:2005mg}%
  \BibitemOpen
  \bibfield  {author} {\bibinfo {author} {\bibfnamefont {Y.}~\bibnamefont
  {Chen}}, \bibinfo {author} {\bibfnamefont {A.}~\bibnamefont {Alexandru}},
  \bibinfo {author} {\bibfnamefont {S.~J.}\ \bibnamefont {Dong}}, \bibinfo
  {author} {\bibfnamefont {T.}~\bibnamefont {Draper}}, \bibinfo {author}
  {\bibfnamefont {I.}~\bibnamefont {Horvath}},  \emph {et~al.},\ }\href
  {\doibase 10.1103/PhysRevD.73.014516} {\bibfield  {journal} {\bibinfo
  {journal} {Phys. Rev. D}\ }\textbf {\bibinfo {volume} {73}},\ \bibinfo
  {pages} {014516} (\bibinfo {year} {2006})},\ \Eprint
  {http://arxiv.org/abs/hep-lat/0510074} {arXiv:hep-lat/0510074
  [hep-lat]}\BibitemShut {NoStop}%
\bibitem [{\citenamefont {Gregory}\ \emph {et~al.}(2012)\citenamefont
  {Gregory}, \citenamefont {Irving}, \citenamefont {Lucini}, \citenamefont
  {McNeile}, \citenamefont {Rago}, \citenamefont {Richards},\ and\
  \citenamefont {Rinaldi}}]{Gregory:2012hu}%
  \BibitemOpen
  \bibfield  {author} {\bibinfo {author} {\bibfnamefont {E.}~\bibnamefont
  {Gregory}}, \bibinfo {author} {\bibfnamefont {A.}~\bibnamefont {Irving}},
  \bibinfo {author} {\bibfnamefont {B.}~\bibnamefont {Lucini}}, \bibinfo
  {author} {\bibfnamefont {C.}~\bibnamefont {McNeile}}, \bibinfo {author}
  {\bibfnamefont {A.}~\bibnamefont {Rago}},  \emph {et~al.},\ }\href {\doibase
  10.1007/JHEP10(2012)170} {\bibfield  {journal} {\bibinfo  {journal} {JHEP}\
  }\textbf {\bibinfo {volume} {10}},\ \bibinfo {pages} {170} (\bibinfo {year}
  {2012})},\ \Eprint {http://arxiv.org/abs/1208.1858} {arXiv:1208.1858
  [hep-lat]}\BibitemShut {NoStop}%
\bibitem [{\citenamefont {Gui}\ \emph {et~al.}(2013)\citenamefont {Gui},
  \citenamefont {Chen}, \citenamefont {Li}, \citenamefont {Liu}, \citenamefont
  {Liu}, \citenamefont {Ma}, \citenamefont {Yang},\ and\ \citenamefont
  {Zhang}}]{Gui:2012gx}%
  \BibitemOpen
  \bibfield  {author} {\bibinfo {author} {\bibfnamefont {L.-C.}\ \bibnamefont
  {Gui}}, \bibinfo {author} {\bibfnamefont {Y.}~\bibnamefont {Chen}}, \bibinfo
  {author} {\bibfnamefont {G.}~\bibnamefont {Li}}, \bibinfo {author}
  {\bibfnamefont {C.}~\bibnamefont {Liu}}, \bibinfo {author} {\bibfnamefont
  {Y.-B.}\ \bibnamefont {Liu}},  \emph {et~al.},\ }\href {\doibase
  10.1103/PhysRevLett.110.021601} {\bibfield  {journal} {\bibinfo  {journal}
  {Phys. Rev. Lett.}\ }\textbf {\bibinfo {volume} {110}},\ \bibinfo {pages}
  {021601} (\bibinfo {year} {2013})}\BibitemShut {NoStop}%
\bibitem [{\citenamefont {Kharzeev}(2021)}]{Kharzeev:2021qkd}%
  \BibitemOpen
  \bibfield  {author} {\bibinfo {author} {\bibfnamefont {D.~E.}\ \bibnamefont
  {Kharzeev}},\ }\href {\doibase 10.1103/PhysRevD.104.054015} {\bibfield
  {journal} {\bibinfo  {journal} {Phys. Rev. D}\ }\textbf {\bibinfo {volume}
  {104}},\ \bibinfo {pages} {054015} (\bibinfo {year} {2021})},\ \Eprint
  {http://arxiv.org/abs/2102.00110} {arXiv:2102.00110 [hep-ph]}\BibitemShut
  {NoStop}%
\bibitem [{\citenamefont {Duran}\ \emph {et~al.}(2023)\citenamefont {Duran},
  \citenamefont {Meziani}, \citenamefont {Joosten}, \citenamefont {Jones},
  \citenamefont {Prasad} \emph {et~al.}}]{Duran:2022xag}%
  \BibitemOpen
  \bibfield  {author} {\bibinfo {author} {\bibfnamefont {B.}~\bibnamefont
  {Duran}}, \bibinfo {author} {\bibfnamefont {Z.~E.}\ \bibnamefont {Meziani}},
  \bibinfo {author} {\bibfnamefont {S.}~\bibnamefont {Joosten}}, \bibinfo
  {author} {\bibfnamefont {M.~K.}\ \bibnamefont {Jones}}, \bibinfo {author}
  {\bibfnamefont {S.}~\bibnamefont {Prasad}},  \emph {et~al.},\ }\href
  {\doibase 10.1038/s41586-023-05730-4} {\bibfield  {journal} {\bibinfo
  {journal} {Nature}\ }\textbf {\bibinfo {volume} {615}},\ \bibinfo {pages}
  {813} (\bibinfo {year} {2023})},\ \Eprint {http://arxiv.org/abs/2207.05212}
  {arXiv:2207.05212 [nucl-ex]}\BibitemShut {NoStop}%
\bibitem [{\citenamefont {Pasquini}\ \emph {et~al.}(2014)\citenamefont
  {Pasquini}, \citenamefont {Polyakov},\ and\ \citenamefont
  {Vanderhaeghen}}]{Pasquini:2014vua}%
  \BibitemOpen
  \bibfield  {author} {\bibinfo {author} {\bibfnamefont {B.}~\bibnamefont
  {Pasquini}}, \bibinfo {author} {\bibfnamefont {M.~V.}\ \bibnamefont
  {Polyakov}}, and \bibinfo {author} {\bibfnamefont {M.}~\bibnamefont
  {Vanderhaeghen}},\ }\href {\doibase 10.1016/j.physletb.2014.10.047}
  {\bibfield  {journal} {\bibinfo  {journal} {Phys. Lett. B}\ }\textbf
  {\bibinfo {volume} {739}},\ \bibinfo {pages} {133} (\bibinfo {year}
  {2014})},\ \Eprint {http://arxiv.org/abs/1407.5960} {arXiv:1407.5960
  [hep-ph]}\BibitemShut {NoStop}%
\bibitem [{\citenamefont {Cao}\ \emph {et~al.}(2025)\citenamefont {Cao},
  \citenamefont {Guo}, \citenamefont {Li},\ and\ \citenamefont
  {Yao}}]{Cao:2024zlf}%
  \BibitemOpen
  \bibfield  {author} {\bibinfo {author} {\bibfnamefont {X.-H.}\ \bibnamefont
  {Cao}}, \bibinfo {author} {\bibfnamefont {F.-K.}\ \bibnamefont {Guo}},
  \bibinfo {author} {\bibfnamefont {Q.-Z.}\ \bibnamefont {Li}}, and \bibinfo
  {author} {\bibfnamefont {D.-L.}\ \bibnamefont {Yao}},\ }\href {\doibase
  10.1038/s41467-025-62278-9} {\bibfield  {journal} {\bibinfo  {journal}
  {Nature Commun.}\ }\textbf {\bibinfo {volume} {16}},\ \bibinfo {pages} {6979}
  (\bibinfo {year} {2025})},\ \Eprint {http://arxiv.org/abs/2411.13398}
  {arXiv:2411.13398 [hep-ph]}\BibitemShut {NoStop}%
\bibitem [{\citenamefont {{BES Collaboration}}(2006)}]{BES:2006ssn}%
  \BibitemOpen
  \bibfield  {author} {\bibinfo {author} {\bibnamefont {{BES Collaboration}}},\
  }\href {\doibase 10.1016/j.physletb.2006.10.004} {\bibfield  {journal}
  {\bibinfo  {journal} {Phys. Lett. B}\ }\textbf {\bibinfo {volume} {642}},\
  \bibinfo {pages} {441} (\bibinfo {year} {2006})},\ \Eprint
  {http://arxiv.org/abs/hep-ex/0603048} {arXiv:hep-ex/0603048
  [hep-ex]}\BibitemShut {NoStop}%
\bibitem [{\citenamefont {Sarantsev}\ \emph {et~al.}(2021)\citenamefont
  {Sarantsev}, \citenamefont {Denisenko}, \citenamefont {Thoma},\ and\
  \citenamefont {Klempt}}]{Sarantsev:2021ein}%
  \BibitemOpen
  \bibfield  {author} {\bibinfo {author} {\bibfnamefont {A.~V.}\ \bibnamefont
  {Sarantsev}}, \bibinfo {author} {\bibfnamefont {I.}~\bibnamefont
  {Denisenko}}, \bibinfo {author} {\bibfnamefont {U.}~\bibnamefont {Thoma}},
  and \bibinfo {author} {\bibfnamefont {E.}~\bibnamefont {Klempt}},\ }\href
  {\doibase 10.1016/j.physletb.2021.136227} {\bibfield  {journal} {\bibinfo
  {journal} {Phys. Lett. B}\ }\textbf {\bibinfo {volume} {816}},\ \bibinfo
  {pages} {136227} (\bibinfo {year} {2021})},\ \Eprint
  {http://arxiv.org/abs/2103.09680} {arXiv:2103.09680 [hep-ph]}\BibitemShut
  {NoStop}%
\bibitem [{\citenamefont {Rodas}\ \emph {et~al.}(2022)\citenamefont {Rodas},
  \citenamefont {Pilloni}, \citenamefont {Albaladejo}, \citenamefont
  {Fern{\'a}ndez-Ram{\'i}rez}, \citenamefont {Mathieu},\ and\ \citenamefont
  {Szczepaniak}}]{Rodas:2022jpsi}%
  \BibitemOpen
  \bibfield  {author} {\bibinfo {author} {\bibfnamefont {A.}~\bibnamefont
  {Rodas}}, \bibinfo {author} {\bibfnamefont {A.}~\bibnamefont {Pilloni}},
  \bibinfo {author} {\bibfnamefont {M.}~\bibnamefont {Albaladejo}}, \bibinfo
  {author} {\bibfnamefont {C.}~\bibnamefont {Fern{\'a}ndez-Ram{\'i}rez}},
  \bibinfo {author} {\bibfnamefont {V.}~\bibnamefont {Mathieu}},  \emph
  {et~al.},\ }\href {\doibase 10.1140/epjc/s10052-022-10014-8} {\bibfield
  {journal} {\bibinfo  {journal} {Eur. Phys. J. C}\ }\textbf {\bibinfo {volume}
  {82}},\ \bibinfo {pages} {80} (\bibinfo {year} {2022})}\BibitemShut {NoStop}%
\bibitem [{\citenamefont {Xiao}\ \emph {et~al.}(2020)\citenamefont {Xiao},
  \citenamefont {Mei{\ss}ner},\ and\ \citenamefont {Oller}}]{Xiao:2019lrj}%
  \BibitemOpen
  \bibfield  {author} {\bibinfo {author} {\bibfnamefont {C.~W.}\ \bibnamefont
  {Xiao}}, \bibinfo {author} {\bibfnamefont {U.-G.}\ \bibnamefont
  {Mei{\ss}ner}}, and \bibinfo {author} {\bibfnamefont {J.~A.}\ \bibnamefont
  {Oller}},\ }\href {\doibase 10.1140/epja/s10050-020-00025-y} {\bibfield
  {journal} {\bibinfo  {journal} {Eur. Phys. J. A}\ }\textbf {\bibinfo {volume}
  {56}},\ \bibinfo {pages} {23} (\bibinfo {year} {2020})},\ \Eprint
  {http://arxiv.org/abs/1907.09072} {arXiv:1907.09072 [hep-ph]}\BibitemShut
  {NoStop}%
\bibitem [{\citenamefont {Sakai}\ \emph {et~al.}(2020)\citenamefont {Sakai},
  \citenamefont {Liang}, \citenamefont {Toledo},\ and\ \citenamefont
  {Oset}}]{Sakai:2019uig}%
  \BibitemOpen
  \bibfield  {author} {\bibinfo {author} {\bibfnamefont {S.}~\bibnamefont
  {Sakai}}, \bibinfo {author} {\bibfnamefont {W.-H.}\ \bibnamefont {Liang}},
  \bibinfo {author} {\bibfnamefont {G.}~\bibnamefont {Toledo}}, and \bibinfo
  {author} {\bibfnamefont {E.}~\bibnamefont {Oset}},\ }\href {\doibase
  10.1103/PhysRevD.101.014005} {\bibfield  {journal} {\bibinfo  {journal}
  {Phys. Rev. D}\ }\textbf {\bibinfo {volume} {101}},\ \bibinfo {pages}
  {014005} (\bibinfo {year} {2020})},\ \Eprint
  {http://arxiv.org/abs/1909.08888} {arXiv:1909.08888 [hep-ph]}\BibitemShut
  {NoStop}%
\bibitem [{\citenamefont {Achasov}\ \emph {et~al.}(2021)\citenamefont
  {Achasov}, \citenamefont {Bennett}, \citenamefont {Kiselev}, \citenamefont
  {Kozyrev},\ and\ \citenamefont {Shestakov}}]{Achasov:2020aun}%
  \BibitemOpen
  \bibfield  {author} {\bibinfo {author} {\bibfnamefont {N.~N.}\ \bibnamefont
  {Achasov}}, \bibinfo {author} {\bibfnamefont {J.~V.}\ \bibnamefont
  {Bennett}}, \bibinfo {author} {\bibfnamefont {A.~V.}\ \bibnamefont
  {Kiselev}}, \bibinfo {author} {\bibfnamefont {E.~A.}\ \bibnamefont
  {Kozyrev}}, and \bibinfo {author} {\bibfnamefont {G.~N.}\ \bibnamefont
  {Shestakov}},\ }\href {\doibase 10.1103/PhysRevD.103.014010} {\bibfield
  {journal} {\bibinfo  {journal} {Phys. Rev. D}\ }\textbf {\bibinfo {volume}
  {103}},\ \bibinfo {pages} {014010} (\bibinfo {year} {2021})},\ \Eprint
  {http://arxiv.org/abs/2009.04191} {arXiv:2009.04191 [hep-ph]}\BibitemShut
  {NoStop}%
\bibitem [{\citenamefont {Danilkin}\ \emph {et~al.}(2026)\citenamefont
  {Danilkin}, \citenamefont {Deineka}, \citenamefont {Passemar},\ and\
  \citenamefont {Vanderhaeghen}}]{Danilkin:2025kyo}%
  \BibitemOpen
  \bibfield  {author} {\bibinfo {author} {\bibfnamefont {I.}~\bibnamefont
  {Danilkin}}, \bibinfo {author} {\bibfnamefont {O.}~\bibnamefont {Deineka}},
  \bibinfo {author} {\bibfnamefont {E.}~\bibnamefont {Passemar}}, and \bibinfo
  {author} {\bibfnamefont {M.}~\bibnamefont {Vanderhaeghen}},\ }\href {\doibase
  10.1016/j.physletb.2026.140327} {\bibfield  {journal} {\bibinfo  {journal}
  {Phys. Lett. B}\ }\textbf {\bibinfo {volume} {875}},\ \bibinfo {pages}
  {140327} (\bibinfo {year} {2026})},\ \Eprint
  {http://arxiv.org/abs/2512.23669} {arXiv:2512.23669 [hep-ph]}\BibitemShut
  {NoStop}%
\bibitem [{\citenamefont {Garc\'ia-Mart\'in}\ and\ \citenamefont
  {Moussallam}(2010)}]{Garcia-Martin:2010kyn}%
  \BibitemOpen
  \bibfield  {author} {\bibinfo {author} {\bibfnamefont {R.}~\bibnamefont
  {Garc\'ia-Mart\'in}} and \bibinfo {author} {\bibfnamefont {B.}~\bibnamefont
  {Moussallam}},\ }\href {\doibase 10.1140/epjc/s10052-010-1471-7} {\bibfield
  {journal} {\bibinfo  {journal} {Eur. Phys. J. C}\ }\textbf {\bibinfo {volume}
  {70}},\ \bibinfo {pages} {155} (\bibinfo {year} {2010})},\ \Eprint
  {http://arxiv.org/abs/1006.5373} {arXiv:1006.5373 [hep-ph]}\BibitemShut
  {NoStop}%
\bibitem [{\citenamefont {Moussallam}(2013)}]{Moussallam:2013una}%
  \BibitemOpen
  \bibfield  {author} {\bibinfo {author} {\bibfnamefont {B.}~\bibnamefont
  {Moussallam}},\ }\href {\doibase 10.1140/epjc/s10052-013-2539-y} {\bibfield
  {journal} {\bibinfo  {journal} {Eur. Phys. J. C}\ }\textbf {\bibinfo {volume}
  {73}},\ \bibinfo {pages} {2539} (\bibinfo {year} {2013})},\ \Eprint
  {http://arxiv.org/abs/1305.3143} {arXiv:1305.3143 [hep-ph]}\BibitemShut
  {NoStop}%
\bibitem [{\citenamefont {Danilkin}\ and\ \citenamefont
  {Vanderhaeghen}(2019)}]{Danilkin:2018qfn}%
  \BibitemOpen
  \bibfield  {author} {\bibinfo {author} {\bibfnamefont {I.}~\bibnamefont
  {Danilkin}} and \bibinfo {author} {\bibfnamefont {M.}~\bibnamefont
  {Vanderhaeghen}},\ }\href {\doibase 10.1016/j.physletb.2018.12.047}
  {\bibfield  {journal} {\bibinfo  {journal} {Phys. Lett. B}\ }\textbf
  {\bibinfo {volume} {789}},\ \bibinfo {pages} {366} (\bibinfo {year}
  {2019})},\ \Eprint {http://arxiv.org/abs/1810.03669} {arXiv:1810.03669
  [hep-ph]}\BibitemShut {NoStop}%
\bibitem [{\citenamefont {Hoferichter}\ and\ \citenamefont
  {Stoffer}(2019)}]{Hoferichter:2019nlq}%
  \BibitemOpen
  \bibfield  {author} {\bibinfo {author} {\bibfnamefont {M.}~\bibnamefont
  {Hoferichter}} and \bibinfo {author} {\bibfnamefont {P.}~\bibnamefont
  {Stoffer}},\ }\href {\doibase 10.1007/JHEP07(2019)073} {\bibfield  {journal}
  {\bibinfo  {journal} {JHEP}\ }\textbf {\bibinfo {volume} {07}},\ \bibinfo
  {pages} {073} (\bibinfo {year} {2019})},\ \Eprint
  {http://arxiv.org/abs/1905.13198} {arXiv:1905.13198 [hep-ph]}\BibitemShut
  {NoStop}%
\bibitem [{\citenamefont {Danilkin}\ \emph {et~al.}(2020)\citenamefont
  {Danilkin}, \citenamefont {Deineka},\ and\ \citenamefont
  {Vanderhaeghen}}]{Danilkin:2019opj}%
  \BibitemOpen
  \bibfield  {author} {\bibinfo {author} {\bibfnamefont {I.}~\bibnamefont
  {Danilkin}}, \bibinfo {author} {\bibfnamefont {O.}~\bibnamefont {Deineka}},
  and \bibinfo {author} {\bibfnamefont {M.}~\bibnamefont {Vanderhaeghen}},\
  }\href {\doibase 10.1103/PhysRevD.101.054008} {\bibfield  {journal} {\bibinfo
   {journal} {Phys. Rev. D}\ }\textbf {\bibinfo {volume} {101}},\ \bibinfo
  {pages} {054008} (\bibinfo {year} {2020})},\ \Eprint
  {http://arxiv.org/abs/1909.04158} {arXiv:1909.04158 [hep-ph]}\BibitemShut
  {NoStop}%
\bibitem [{\citenamefont {Colangelo}\ \emph
  {et~al.}(2017{\natexlab{a}})\citenamefont {Colangelo}, \citenamefont
  {Hoferichter}, \citenamefont {Procura},\ and\ \citenamefont
  {Stoffer}}]{Colangelo:2017fiz}%
  \BibitemOpen
  \bibfield  {author} {\bibinfo {author} {\bibfnamefont {G.}~\bibnamefont
  {Colangelo}}, \bibinfo {author} {\bibfnamefont {M.}~\bibnamefont
  {Hoferichter}}, \bibinfo {author} {\bibfnamefont {M.}~\bibnamefont
  {Procura}}, and \bibinfo {author} {\bibfnamefont {P.}~\bibnamefont
  {Stoffer}},\ }\href {\doibase 10.1007/JHEP04(2017)161} {\bibfield  {journal}
  {\bibinfo  {journal} {JHEP}\ }\textbf {\bibinfo {volume} {04}},\ \bibinfo
  {pages} {161} (\bibinfo {year} {2017}{\natexlab{a}})},\ \Eprint
  {http://arxiv.org/abs/1702.07347} {arXiv:1702.07347 [hep-ph]}\BibitemShut
  {NoStop}%
\bibitem [{\citenamefont {Colangelo}\ \emph
  {et~al.}(2017{\natexlab{b}})\citenamefont {Colangelo}, \citenamefont
  {Hoferichter}, \citenamefont {Procura},\ and\ \citenamefont
  {Stoffer}}]{Colangelo:2017qdm}%
  \BibitemOpen
  \bibfield  {author} {\bibinfo {author} {\bibfnamefont {G.}~\bibnamefont
  {Colangelo}}, \bibinfo {author} {\bibfnamefont {M.}~\bibnamefont
  {Hoferichter}}, \bibinfo {author} {\bibfnamefont {M.}~\bibnamefont
  {Procura}}, and \bibinfo {author} {\bibfnamefont {P.}~\bibnamefont
  {Stoffer}},\ }\href {\doibase 10.1103/PhysRevLett.118.232001} {\bibfield
  {journal} {\bibinfo  {journal} {Phys. Rev. Lett.}\ }\textbf {\bibinfo
  {volume} {118}},\ \bibinfo {pages} {232001} (\bibinfo {year}
  {2017}{\natexlab{b}})},\ \Eprint {http://arxiv.org/abs/1701.06554}
  {arXiv:1701.06554 [hep-ph]}\BibitemShut {NoStop}%
\bibitem [{\citenamefont {Danilkin}\ \emph
  {et~al.}(2021{\natexlab{a}})\citenamefont {Danilkin}, \citenamefont
  {Hoferichter},\ and\ \citenamefont {Stoffer}}]{Danilkin:2021icn}%
  \BibitemOpen
  \bibfield  {author} {\bibinfo {author} {\bibfnamefont {I.}~\bibnamefont
  {Danilkin}}, \bibinfo {author} {\bibfnamefont {M.}~\bibnamefont
  {Hoferichter}}, and \bibinfo {author} {\bibfnamefont {P.}~\bibnamefont
  {Stoffer}},\ }\href {\doibase 10.1016/j.physletb.2021.136502} {\bibfield
  {journal} {\bibinfo  {journal} {Phys. Lett. B}\ }\textbf {\bibinfo {volume}
  {820}},\ \bibinfo {pages} {136502} (\bibinfo {year} {2021}{\natexlab{a}})},\
  \Eprint {http://arxiv.org/abs/2105.01666} {arXiv:2105.01666
  [hep-ph]}\BibitemShut {NoStop}%
\bibitem [{\citenamefont {Hoid}\ \emph {et~al.}(2026)\citenamefont {Hoid},
  \citenamefont {Danilkin},\ and\ \citenamefont
  {Vanderhaeghen}}]{Hoid:2026atr}%
  \BibitemOpen
  \bibfield  {author} {\bibinfo {author} {\bibfnamefont {B.-L.}\ \bibnamefont
  {Hoid}}, \bibinfo {author} {\bibfnamefont {I.}~\bibnamefont {Danilkin}}, and
  \bibinfo {author} {\bibfnamefont {M.}~\bibnamefont {Vanderhaeghen}},\ }\href
  {\doibase 10.1103/dlcc-jm4d} {\bibfield  {journal} {\bibinfo  {journal}
  {Phys. Rev. D}\ }\textbf {\bibinfo {volume} {113}},\ \bibinfo {pages}
  {114046} (\bibinfo {year} {2026})},\ \Eprint
  {http://arxiv.org/abs/2602.15100} {arXiv:2602.15100 [hep-ph]}\BibitemShut
  {NoStop}%
\bibitem [{\citenamefont {Moussallam}(2021)}]{Moussallam:2021dpk}%
  \BibitemOpen
  \bibfield  {author} {\bibinfo {author} {\bibfnamefont {B.}~\bibnamefont
  {Moussallam}},\ }\href {\doibase 10.1140/epjc/s10052-021-09772-8} {\bibfield
  {journal} {\bibinfo  {journal} {Eur. Phys. J. C}\ }\textbf {\bibinfo {volume}
  {81}},\ \bibinfo {pages} {993} (\bibinfo {year} {2021})},\ \Eprint
  {http://arxiv.org/abs/2107.14147} {arXiv:2107.14147 [hep-ph]}\BibitemShut
  {NoStop}%
\bibitem [{\citenamefont {Bardeen}\ and\ \citenamefont
  {Tung}(1968)}]{Bardeen:1968ebo}%
  \BibitemOpen
  \bibfield  {author} {\bibinfo {author} {\bibfnamefont {W.~A.}\ \bibnamefont
  {Bardeen}} and \bibinfo {author} {\bibfnamefont {W.~K.}\ \bibnamefont
  {Tung}},\ }\href {\doibase 10.1103/PhysRev.173.1423} {\bibfield  {journal}
  {\bibinfo  {journal} {Phys. Rev.}\ }\textbf {\bibinfo {volume} {173}},\
  \bibinfo {pages} {1423} (\bibinfo {year} {1968})},\ \bibinfo {note}
  {[Erratum: Phys. Rev. D {\bf 4}, 3229 (1971)]}\BibitemShut {NoStop}%
\bibitem [{\citenamefont {Tarrach}(1975)}]{Tarrach:1975tu}%
  \BibitemOpen
  \bibfield  {author} {\bibinfo {author} {\bibfnamefont {R.}~\bibnamefont
  {Tarrach}},\ }\href {\doibase 10.1007/BF02894857} {\bibfield  {journal}
  {\bibinfo  {journal} {Nuovo Cim. A}\ }\textbf {\bibinfo {volume} {28}},\
  \bibinfo {pages} {409} (\bibinfo {year} {1975})}\BibitemShut {NoStop}%
\bibitem [{\citenamefont {Drechsel}\ \emph {et~al.}(1998)\citenamefont
  {Drechsel}, \citenamefont {Kn{\"o}chlein}, \citenamefont {Korchin},
  \citenamefont {Metz},\ and\ \citenamefont {Scherer}}]{Drechsel:1997xv}%
  \BibitemOpen
  \bibfield  {author} {\bibinfo {author} {\bibfnamefont {D.}~\bibnamefont
  {Drechsel}}, \bibinfo {author} {\bibfnamefont {G.}~\bibnamefont
  {Kn{\"o}chlein}}, \bibinfo {author} {\bibfnamefont {A.~Y.}\ \bibnamefont
  {Korchin}}, \bibinfo {author} {\bibfnamefont {A.}~\bibnamefont {Metz}}, and
  \bibinfo {author} {\bibfnamefont {S.}~\bibnamefont {Scherer}},\ }\href
  {\doibase 10.1103/PhysRevC.57.941} {\bibfield  {journal} {\bibinfo  {journal}
  {Phys. Rev. C}\ }\textbf {\bibinfo {volume} {57}},\ \bibinfo {pages} {941}
  (\bibinfo {year} {1998})},\ \Eprint {http://arxiv.org/abs/nucl-th/9704064}
  {arXiv:nucl-th/9704064}\BibitemShut {NoStop}%
\bibitem [{\citenamefont {Chen}\ \emph {et~al.}(2015)\citenamefont {Chen},
  \citenamefont {Guo},\ and\ \citenamefont {Zou}}]{Chen:2014yta}%
  \BibitemOpen
  \bibfield  {author} {\bibinfo {author} {\bibfnamefont {Y.-H.}\ \bibnamefont
  {Chen}}, \bibinfo {author} {\bibfnamefont {Z.-H.}\ \bibnamefont {Guo}}, and
  \bibinfo {author} {\bibfnamefont {B.-S.}\ \bibnamefont {Zou}},\ }\href
  {\doibase 10.1103/PhysRevD.91.014010} {\bibfield  {journal} {\bibinfo
  {journal} {Phys. Rev. D}\ }\textbf {\bibinfo {volume} {91}},\ \bibinfo
  {pages} {014010} (\bibinfo {year} {2015})},\ \Eprint
  {http://arxiv.org/abs/1411.1159} {arXiv:1411.1159 [hep-ph]}\BibitemShut
  {NoStop}%
\bibitem [{\citenamefont {Baldini~Ferroli}\ \emph {et~al.}(2018)\citenamefont
  {Baldini~Ferroli}, \citenamefont {Mangoni},\ and\ \citenamefont
  {Pacetti}}]{BaldiniFerroli:2016mbs}%
  \BibitemOpen
  \bibfield  {author} {\bibinfo {author} {\bibfnamefont {R.}~\bibnamefont
  {Baldini~Ferroli}}, \bibinfo {author} {\bibfnamefont {A.}~\bibnamefont
  {Mangoni}}, and \bibinfo {author} {\bibfnamefont {S.}~\bibnamefont
  {Pacetti}},\ }\href {\doibase 10.1103/PhysRevC.98.045210} {\bibfield
  {journal} {\bibinfo  {journal} {Phys. Rev. C}\ }\textbf {\bibinfo {volume}
  {98}},\ \bibinfo {pages} {045210} (\bibinfo {year} {2018})},\ \Eprint
  {http://arxiv.org/abs/1611.04437} {arXiv:1611.04437 [hep-ph]}\BibitemShut
  {NoStop}%
\bibitem [{\citenamefont {Cao}\ \emph {et~al.}(2026)\citenamefont {Cao},
  \citenamefont {Guo}, \citenamefont {Hanhart},\ and\ \citenamefont
  {Kubis}}]{Cao:2025ncx}%
  \BibitemOpen
  \bibfield  {author} {\bibinfo {author} {\bibfnamefont {X.-H.}\ \bibnamefont
  {Cao}}, \bibinfo {author} {\bibfnamefont {F.-K.}\ \bibnamefont {Guo}},
  \bibinfo {author} {\bibfnamefont {C.}~\bibnamefont {Hanhart}}, and \bibinfo
  {author} {\bibfnamefont {B.}~\bibnamefont {Kubis}},\ }\href {\doibase
  10.1103/gmj6-hd9j} {\bibfield  {journal} {\bibinfo  {journal} {Phys. Rev. D}\
  }\textbf {\bibinfo {volume} {113}},\ \bibinfo {pages} {074030} (\bibinfo
  {year} {2026})},\ \Eprint {http://arxiv.org/abs/2512.00501} {arXiv:2512.00501
  [hep-ph]}\BibitemShut {NoStop}%
\bibitem [{\citenamefont {Jacob}\ and\ \citenamefont
  {Wick}(1959)}]{Jacob:1959at}%
  \BibitemOpen
  \bibfield  {author} {\bibinfo {author} {\bibfnamefont {M.}~\bibnamefont
  {Jacob}} and \bibinfo {author} {\bibfnamefont {G.~C.}\ \bibnamefont {Wick}},\
  }\href {\doibase 10.1016/0003-4916(59)90051-X} {\bibfield  {journal}
  {\bibinfo  {journal} {Annals Phys.}\ }\textbf {\bibinfo {volume} {7}},\
  \bibinfo {pages} {404} (\bibinfo {year} {1959})},\ \bibinfo {note} {[Annals
  Phys. {\bf 281}, 774 (2000)]}\BibitemShut {NoStop}%
\bibitem [{\citenamefont {Low}(1958)}]{Low:1958sn}%
  \BibitemOpen
  \bibfield  {author} {\bibinfo {author} {\bibfnamefont {F.~E.}\ \bibnamefont
  {Low}},\ }\href {\doibase 10.1103/PhysRev.110.974} {\bibfield  {journal}
  {\bibinfo  {journal} {Phys. Rev.}\ }\textbf {\bibinfo {volume} {110}},\
  \bibinfo {pages} {974} (\bibinfo {year} {1958})}\BibitemShut {NoStop}%
\bibitem [{\citenamefont {Karl}\ \emph {et~al.}(1976)\citenamefont {Karl},
  \citenamefont {Meshkov},\ and\ \citenamefont {Rosner}}]{Karl:1975qf}%
  \BibitemOpen
  \bibfield  {author} {\bibinfo {author} {\bibfnamefont {G.}~\bibnamefont
  {Karl}}, \bibinfo {author} {\bibfnamefont {S.}~\bibnamefont {Meshkov}}, and
  \bibinfo {author} {\bibfnamefont {J.~L.}\ \bibnamefont {Rosner}},\ }\href
  {\doibase 10.1103/PhysRevD.13.1203} {\bibfield  {journal} {\bibinfo
  {journal} {Phys. Rev. D}\ }\textbf {\bibinfo {volume} {13}},\ \bibinfo
  {pages} {1203} (\bibinfo {year} {1976})}\BibitemShut {NoStop}%
\bibitem [{\citenamefont {Olsson}\ and\ \citenamefont
  {Suchyta}(1986)}]{Olsson:1986dn}%
  \BibitemOpen
  \bibfield  {author} {\bibinfo {author} {\bibfnamefont {M.~G.}\ \bibnamefont
  {Olsson}} and \bibinfo {author} {\bibfnamefont {C.~J.}\ \bibnamefont
  {Suchyta}, \bibfnamefont {III}},\ }\href {\doibase 10.1103/PhysRevD.34.2043}
  {\bibfield  {journal} {\bibinfo  {journal} {Phys. Rev. D}\ }\textbf {\bibinfo
  {volume} {34}},\ \bibinfo {pages} {2043} (\bibinfo {year}
  {1986})}\BibitemShut {NoStop}%
\bibitem [{\citenamefont {Sebastian}\ \emph {et~al.}(1992)\citenamefont
  {Sebastian}, \citenamefont {Grotch},\ and\ \citenamefont
  {Ridener}}]{Sebastian:1992qe}%
  \BibitemOpen
  \bibfield  {author} {\bibinfo {author} {\bibfnamefont {K.~J.}\ \bibnamefont
  {Sebastian}}, \bibinfo {author} {\bibfnamefont {H.}~\bibnamefont {Grotch}},
  and \bibinfo {author} {\bibfnamefont {F.~L.}\ \bibnamefont {Ridener},
  \bibfnamefont {Jr.}},\ }\href {\doibase 10.1103/PhysRevD.45.3163} {\bibfield
  {journal} {\bibinfo  {journal} {Phys. Rev. D}\ }\textbf {\bibinfo {volume}
  {45}},\ \bibinfo {pages} {3163} (\bibinfo {year} {1992})}\BibitemShut
  {NoStop}%
\bibitem [{\citenamefont {Danilkin}\ \emph
  {et~al.}(2021{\natexlab{b}})\citenamefont {Danilkin}, \citenamefont
  {Deineka},\ and\ \citenamefont {Vanderhaeghen}}]{Danilkin:2020pak}%
  \BibitemOpen
  \bibfield  {author} {\bibinfo {author} {\bibfnamefont {I.}~\bibnamefont
  {Danilkin}}, \bibinfo {author} {\bibfnamefont {O.}~\bibnamefont {Deineka}},
  and \bibinfo {author} {\bibfnamefont {M.}~\bibnamefont {Vanderhaeghen}},\
  }\href {\doibase 10.1103/PhysRevD.103.114023} {\bibfield  {journal} {\bibinfo
   {journal} {Phys. Rev. D}\ }\textbf {\bibinfo {volume} {103}},\ \bibinfo
  {pages} {114023} (\bibinfo {year} {2021}{\natexlab{b}})},\ \Eprint
  {http://arxiv.org/abs/2012.11636} {arXiv:2012.11636 [hep-ph]}\BibitemShut
  {NoStop}%
\bibitem [{\citenamefont {Garc\'ia-Mart\'in}\ \emph {et~al.}(2011)\citenamefont
  {Garc\'ia-Mart\'in}, \citenamefont {Kami\'nski}, \citenamefont {Pel\'aez},
  \citenamefont {Ruiz~de Elvira},\ and\ \citenamefont
  {Yndur\'ain}}]{Garcia-Martin:2011iqs}%
  \BibitemOpen
  \bibfield  {author} {\bibinfo {author} {\bibfnamefont {R.}~\bibnamefont
  {Garc\'ia-Mart\'in}}, \bibinfo {author} {\bibfnamefont {R.}~\bibnamefont
  {Kami\'nski}}, \bibinfo {author} {\bibfnamefont {J.~R.}\ \bibnamefont
  {Pel\'aez}}, \bibinfo {author} {\bibfnamefont {J.}~\bibnamefont {Ruiz~de
  Elvira}}, and \bibinfo {author} {\bibfnamefont {F.~J.}\ \bibnamefont
  {Yndur\'ain}},\ }\href {\doibase 10.1103/PhysRevD.83.074004} {\bibfield
  {journal} {\bibinfo  {journal} {Phys. Rev. D}\ }\textbf {\bibinfo {volume}
  {83}},\ \bibinfo {pages} {074004} (\bibinfo {year} {2011})},\ \Eprint
  {http://arxiv.org/abs/1102.2183} {arXiv:1102.2183 [hep-ph]}\BibitemShut
  {NoStop}%
\bibitem [{\citenamefont {Pel{\'a}ez}\ and\ \citenamefont
  {Rodas}(2022)}]{Pelaez:2020gnd}%
  \BibitemOpen
  \bibfield  {author} {\bibinfo {author} {\bibfnamefont {J.~R.}\ \bibnamefont
  {Pel{\'a}ez}} and \bibinfo {author} {\bibfnamefont {A.}~\bibnamefont
  {Rodas}},\ }\href {\doibase 10.1016/j.physrep.2022.03.004} {\bibfield
  {journal} {\bibinfo  {journal} {Phys. Rept.}\ }\textbf {\bibinfo {volume}
  {969}},\ \bibinfo {pages} {1} (\bibinfo {year} {2022})},\ \Eprint
  {http://arxiv.org/abs/2010.11222} {arXiv:2010.11222 [hep-ph]}\BibitemShut
  {NoStop}%
\bibitem [{\citenamefont {Niecknig}\ \emph {et~al.}(2012)\citenamefont
  {Niecknig}, \citenamefont {Kubis},\ and\ \citenamefont
  {Schneider}}]{Niecknig:2012sj}%
  \BibitemOpen
  \bibfield  {author} {\bibinfo {author} {\bibfnamefont {F.}~\bibnamefont
  {Niecknig}}, \bibinfo {author} {\bibfnamefont {B.}~\bibnamefont {Kubis}}, and
  \bibinfo {author} {\bibfnamefont {S.~P.}\ \bibnamefont {Schneider}},\ }\href
  {\doibase 10.1140/epjc/s10052-012-2014-1} {\bibfield  {journal} {\bibinfo
  {journal} {Eur. Phys. J. C}\ }\textbf {\bibinfo {volume} {72}},\ \bibinfo
  {pages} {2014} (\bibinfo {year} {2012})},\ \Eprint
  {http://arxiv.org/abs/1203.2501} {arXiv:1203.2501 [hep-ph]}\BibitemShut
  {NoStop}%
\bibitem [{\citenamefont {Schneider}\ \emph {et~al.}(2012)\citenamefont
  {Schneider}, \citenamefont {Kubis},\ and\ \citenamefont
  {Niecknig}}]{Schneider:2012ez}%
  \BibitemOpen
  \bibfield  {author} {\bibinfo {author} {\bibfnamefont {S.~P.}\ \bibnamefont
  {Schneider}}, \bibinfo {author} {\bibfnamefont {B.}~\bibnamefont {Kubis}},
  and \bibinfo {author} {\bibfnamefont {F.}~\bibnamefont {Niecknig}},\ }\href
  {\doibase 10.1103/PhysRevD.86.054013} {\bibfield  {journal} {\bibinfo
  {journal} {Phys. Rev. D}\ }\textbf {\bibinfo {volume} {86}},\ \bibinfo
  {pages} {054013} (\bibinfo {year} {2012})},\ \Eprint
  {http://arxiv.org/abs/1206.3098} {arXiv:1206.3098 [hep-ph]}\BibitemShut
  {NoStop}%
\bibitem [{\citenamefont {Danilkin}\ \emph {et~al.}(2015)\citenamefont
  {Danilkin} \emph {et~al.}}]{Danilkin:2014cra}%
  \BibitemOpen
  \bibfield  {author} {\bibinfo {author} {\bibfnamefont {I.~V.}\ \bibnamefont
  {Danilkin}}  \emph {et~al.},\ }\href {\doibase 10.1103/PhysRevD.91.094029}
  {\bibfield  {journal} {\bibinfo  {journal} {Phys. Rev. D}\ }\textbf {\bibinfo
  {volume} {91}},\ \bibinfo {pages} {094029} (\bibinfo {year} {2015})},\
  \Eprint {http://arxiv.org/abs/1409.7708} {arXiv:1409.7708
  [hep-ph]}\BibitemShut {NoStop}%
\bibitem [{\citenamefont {Albaladejo}\ \emph {et~al.}(2020)\citenamefont
  {Albaladejo}, \citenamefont {Danilkin}, \citenamefont
  {Gonz\`alez-Sol\'\i{}s}, \citenamefont {Winney}, \citenamefont
  {Fern{\'a}ndez-Ram{\'i}rez}, \citenamefont {Hiller~Blin}, \citenamefont
  {Mathieu}, \citenamefont {Mikhasenko}, \citenamefont {Pilloni},\ and\
  \citenamefont {Szczepaniak}}]{JPAC:2020umo}%
  \BibitemOpen
  \bibfield  {author} {\bibinfo {author} {\bibfnamefont {M.}~\bibnamefont
  {Albaladejo}}, \bibinfo {author} {\bibfnamefont {I.}~\bibnamefont
  {Danilkin}}, \bibinfo {author} {\bibfnamefont {S.}~\bibnamefont
  {Gonz\`alez-Sol\'\i{}s}}, \bibinfo {author} {\bibfnamefont {D.}~\bibnamefont
  {Winney}}, \bibinfo {author} {\bibfnamefont {C.}~\bibnamefont
  {Fern{\'a}ndez-Ram{\'i}rez}},  \emph {et~al.} (\bibinfo {collaboration}
  {JPAC}),\ }\href {\doibase 10.1140/epjc/s10052-020-08576-6} {\bibfield
  {journal} {\bibinfo  {journal} {Eur. Phys. J. C}\ }\textbf {\bibinfo {volume}
  {80}},\ \bibinfo {pages} {1107} (\bibinfo {year} {2020})},\ \Eprint
  {http://arxiv.org/abs/2006.01058} {arXiv:2006.01058 [hep-ph]}\BibitemShut
  {NoStop}%
\bibitem [{\citenamefont {Pel{\'a}ez}\ \emph {et~al.}(2025)\citenamefont
  {Pel{\'a}ez}, \citenamefont {Rab{\'a}n},\ and\ \citenamefont {Ruiz~de
  Elvira}}]{Pelaez:2025gpp}%
  \BibitemOpen
  \bibfield  {author} {\bibinfo {author} {\bibfnamefont {J.~R.}\ \bibnamefont
  {Pel{\'a}ez}}, \bibinfo {author} {\bibfnamefont {P.}~\bibnamefont
  {Rab{\'a}n}}, and \bibinfo {author} {\bibfnamefont {J.}~\bibnamefont {Ruiz~de
  Elvira}},\ }\href {\doibase 10.1103/PhysRevD.111.074003} {\bibfield
  {journal} {\bibinfo  {journal} {Phys. Rev. D}\ }\textbf {\bibinfo {volume}
  {111}},\ \bibinfo {pages} {074003} (\bibinfo {year} {2025})}\BibitemShut
  {NoStop}%
\bibitem [{\citenamefont {D'Agostini}(1994)}]{DAgostini:1993arp}%
  \BibitemOpen
  \bibfield  {author} {\bibinfo {author} {\bibfnamefont {G.}~\bibnamefont
  {D'Agostini}},\ }\href {\doibase 10.1016/0168-9002(94)90719-6} {\bibfield
  {journal} {\bibinfo  {journal} {Nucl. Instrum. Meth. A}\ }\textbf {\bibinfo
  {volume} {346}},\ \bibinfo {pages} {306} (\bibinfo {year}
  {1994})}\BibitemShut {NoStop}%
\bibitem [{\citenamefont {Ball}\ \emph {et~al.}(2010)\citenamefont {Ball},
  \citenamefont {Del~Debbio}, \citenamefont {Forte}, \citenamefont {Guffanti},
  \citenamefont {Latorre}, \citenamefont {Rojo},\ and\ \citenamefont
  {Ubiali}}]{Ball:2009qv}%
  \BibitemOpen
  \bibfield  {author} {\bibinfo {author} {\bibfnamefont {R.~D.}\ \bibnamefont
  {Ball}}, \bibinfo {author} {\bibfnamefont {L.}~\bibnamefont {Del~Debbio}},
  \bibinfo {author} {\bibfnamefont {S.}~\bibnamefont {Forte}}, \bibinfo
  {author} {\bibfnamefont {A.}~\bibnamefont {Guffanti}}, \bibinfo {author}
  {\bibfnamefont {J.~I.}\ \bibnamefont {Latorre}},  \emph {et~al.} (\bibinfo
  {collaboration} {NNPDF}),\ }\href {\doibase 10.1007/JHEP05(2010)075}
  {\bibfield  {journal} {\bibinfo  {journal} {JHEP}\ }\textbf {\bibinfo
  {volume} {05}},\ \bibinfo {pages} {075} (\bibinfo {year} {2010})},\ \Eprint
  {http://arxiv.org/abs/0912.2276} {arXiv:0912.2276 [hep-ph]}\BibitemShut
  {NoStop}%
\bibitem [{\citenamefont {Hoferichter}\ \emph {et~al.}(2019)\citenamefont
  {Hoferichter}, \citenamefont {Hoid},\ and\ \citenamefont
  {Kubis}}]{Hoferichter:2019gzf}%
  \BibitemOpen
  \bibfield  {author} {\bibinfo {author} {\bibfnamefont {M.}~\bibnamefont
  {Hoferichter}}, \bibinfo {author} {\bibfnamefont {B.-L.}\ \bibnamefont
  {Hoid}}, and \bibinfo {author} {\bibfnamefont {B.}~\bibnamefont {Kubis}},\
  }\href {\doibase 10.1007/JHEP08(2019)137} {\bibfield  {journal} {\bibinfo
  {journal} {JHEP}\ }\textbf {\bibinfo {volume} {08}},\ \bibinfo {pages} {137}
  (\bibinfo {year} {2019})},\ \Eprint {http://arxiv.org/abs/1907.01556}
  {arXiv:1907.01556 [hep-ph]}\BibitemShut {NoStop}%
\bibitem [{\citenamefont {Hoid}\ \emph {et~al.}(2020)\citenamefont {Hoid},
  \citenamefont {Hoferichter},\ and\ \citenamefont {Kubis}}]{Hoid:2020xjs}%
  \BibitemOpen
  \bibfield  {author} {\bibinfo {author} {\bibfnamefont {B.-L.}\ \bibnamefont
  {Hoid}}, \bibinfo {author} {\bibfnamefont {M.}~\bibnamefont {Hoferichter}},
  and \bibinfo {author} {\bibfnamefont {B.}~\bibnamefont {Kubis}},\ }\href
  {\doibase 10.1140/epjc/s10052-020-08550-2} {\bibfield  {journal} {\bibinfo
  {journal} {Eur. Phys. J. C}\ }\textbf {\bibinfo {volume} {80}},\ \bibinfo
  {pages} {988} (\bibinfo {year} {2020})},\ \Eprint
  {http://arxiv.org/abs/2007.12696} {arXiv:2007.12696 [hep-ph]}\BibitemShut
  {NoStop}%
\bibitem [{\citenamefont {Barrelet}(1972)}]{Barrelet:1971pw}%
  \BibitemOpen
  \bibfield  {author} {\bibinfo {author} {\bibfnamefont {E.}~\bibnamefont
  {Barrelet}},\ }\href {\doibase 10.1007/BF02732655} {\bibfield  {journal}
  {\bibinfo  {journal} {Nuovo Cim. A}\ }\textbf {\bibinfo {volume} {8}},\
  \bibinfo {pages} {331} (\bibinfo {year} {1972})}\BibitemShut {NoStop}%
\bibitem [{\citenamefont {Navas}\ \emph {et~al.}(2024)\citenamefont {Navas}
  \emph {et~al.}}]{ParticleDataGroup:2024cfk}%
  \BibitemOpen
  \bibfield  {author} {\bibinfo {author} {\bibfnamefont {S.}~\bibnamefont
  {Navas}}  \emph {et~al.} (\bibinfo {collaboration} {Particle Data Group}),\
  }\href {\doibase 10.1103/PhysRevD.110.030001} {\bibfield  {journal} {\bibinfo
   {journal} {Phys. Rev. D}\ }\textbf {\bibinfo {volume} {110}},\ \bibinfo
  {pages} {030001} (\bibinfo {year} {2024})}\BibitemShut {NoStop}%
\bibitem [{\citenamefont {Hoferichter}\ \emph {et~al.}(2025)\citenamefont
  {Hoferichter}, \citenamefont {Hoid},\ and\ \citenamefont
  {Kubis}}]{Hoferichter:2025lcz}%
  \BibitemOpen
  \bibfield  {author} {\bibinfo {author} {\bibfnamefont {M.}~\bibnamefont
  {Hoferichter}}, \bibinfo {author} {\bibfnamefont {B.-L.}\ \bibnamefont
  {Hoid}}, and \bibinfo {author} {\bibfnamefont {B.}~\bibnamefont {Kubis}},\
  }\href {\doibase 10.1007/JHEP07(2025)095} {\bibfield  {journal} {\bibinfo
  {journal} {JHEP}\ }\textbf {\bibinfo {volume} {07}},\ \bibinfo {pages} {095}
  (\bibinfo {year} {2025})},\ \Eprint {http://arxiv.org/abs/2504.13827}
  {arXiv:2504.13827 [hep-ph]}\BibitemShut {NoStop}%
\bibitem [{\citenamefont {Oller}(2019)}]{Oller:2019rej}%
  \BibitemOpen
  \bibfield  {author} {\bibinfo {author} {\bibfnamefont {J.~A.}\ \bibnamefont
  {Oller}},\ }\href {\doibase 10.1007/978-3-030-13582-9} {\emph {\bibinfo
  {title} {{A Brief Introduction to Dispersion Relations}}}},\ SpringerBriefs
  in Physics\ (\bibinfo  {publisher} {Springer},\ \bibinfo {year}
  {2019})\BibitemShut {NoStop}%
\bibitem [{\citenamefont {Zanke}\ \emph {et~al.}(2021)\citenamefont {Zanke},
  \citenamefont {Hoferichter},\ and\ \citenamefont {Kubis}}]{Zanke:2021wiq}%
  \BibitemOpen
  \bibfield  {author} {\bibinfo {author} {\bibfnamefont {M.}~\bibnamefont
  {Zanke}}, \bibinfo {author} {\bibfnamefont {M.}~\bibnamefont {Hoferichter}},
  and \bibinfo {author} {\bibfnamefont {B.}~\bibnamefont {Kubis}},\ }\href
  {\doibase 10.1007/JHEP07(2021)106} {\bibfield  {journal} {\bibinfo  {journal}
  {JHEP}\ }\textbf {\bibinfo {volume} {07}},\ \bibinfo {pages} {106} (\bibinfo
  {year} {2021})},\ \Eprint {http://arxiv.org/abs/2103.09829} {arXiv:2103.09829
  [hep-ph]}\BibitemShut {NoStop}%
\bibitem [{\citenamefont {Messerli}\ \emph {et~al.}(2026)\citenamefont
  {Messerli}, \citenamefont {Hoferichter}, \citenamefont {Hoid}, \citenamefont
  {Holz},\ and\ \citenamefont {Kubis}}]{Messerli:2025rnv}%
  \BibitemOpen
  \bibfield  {author} {\bibinfo {author} {\bibfnamefont {N.}~\bibnamefont
  {Messerli}}, \bibinfo {author} {\bibfnamefont {M.}~\bibnamefont
  {Hoferichter}}, \bibinfo {author} {\bibfnamefont {B.-L.}\ \bibnamefont
  {Hoid}}, \bibinfo {author} {\bibfnamefont {S.}~\bibnamefont {Holz}}, and
  \bibinfo {author} {\bibfnamefont {B.}~\bibnamefont {Kubis}},\ }\href
  {\doibase 10.1007/JHEP04(2026)088} {\bibfield  {journal} {\bibinfo  {journal}
  {JHEP}\ }\textbf {\bibinfo {volume} {04}},\ \bibinfo {pages} {088} (\bibinfo
  {year} {2026})},\ \Eprint {http://arxiv.org/abs/2512.13776} {arXiv:2512.13776
  [hep-ph]}\BibitemShut {NoStop}%
\end{thebibliography}%

\end{document}